\documentclass[review]{elsarticle}

\usepackage{lineno}
\modulolinenumbers[5]

\usepackage{color}
\usepackage{array}
\usepackage{amssymb}
\usepackage[colorlinks,urlcolor=blue]{hyperref}
\usepackage{amsmath}
\usepackage[export]{adjustbox}

\usepackage[margin=2.5cm]{geometry} 

\journal{}

\biboptions{authoryear}

\begin{document} 

\begin{frontmatter}

\title{A reduced-order, rotation-based model for thin hard-magnetic plates}



\author[mymainaddress]{Dong Yan}
\ead{dong.yan@epfl.ch}

\author[mymainaddress]{Bastien F. G. Aymon}
\ead{bastien.aymon@epfl.ch}

\author[mymainaddress]{Pedro M. Reis\corref{mycorrespondingauthor}}
\cortext[mycorrespondingauthor]{Corresponding author}
\ead{pedro.reis@epfl.ch}

\address[mymainaddress]{Flexible Structures Laboratory, Institute of Mechanical Engineering,\\
	\'{E}cole Polytechnique F\'{e}d\'{e}rale de Lausanne,
	1015 Lausanne, Switzerland\\
    }

\begin{abstract}

We develop a reduced-order model for thin plates made of hard magnetorheological elastomers (hard-MREs), which are materials composed of hard-magnetic particles embedded in a polymeric matrix. First, we propose a new magnetic potential, as an alternative to an existing torque-based 3D continuum theory of hard-MREs, obtained by reformulating the remnant magnetization of a deformed hard-MRE body. Specifically, the magnetizations in the initial and current configurations are related by the rotation tensor decomposed from the deformation gradient, independently of stretching deformation. This description is motivated by recently reported observations in microscopic homogenization simulations. Then, we derive a 2D plate model through the dimensional reduction of our proposed rotation-based 3D theory. For comparison, we also provide a second plate model derived from the existing 3D theory.  Finally, we perform precision experiments to thoroughly evaluate the proposed 3D and 2D models on hard-magnetic plates under various magnetic and mechanical loading conditions. We demonstrate that our rotation-based modification of the magnetic potential is crucial in correctly capturing the behavior of plates subjected to an applied field aligned with the magnetization, and undergoing in-plane stretching. In all the tested cases, our rotation-based 3D and 2D models yield predictions in excellent quantitative agreement with the experiments and can thus serve as predictive tools for the rational design of hard-magnetic plate structures.

\end{abstract}

\begin{keyword}
Thin plates \sep Hard magnetorheological elastomers \sep Reduced-order model \sep Magnetic potential \sep Magnetization \sep Rotation tensor
\end{keyword}

\end{frontmatter}

\clearpage


\section{Introduction}
\label{sec:introduction}

Soft active materials are compliant to mechanical deformations and responsive to external stimuli, making them ideal candidates for smart structures and systems in applications ranging from sensors and actuators to tissue engineering and energy harvesting~\citep{kim_magnetic_2022,wang_multi-functional_2022}. This family of materials includes liquid crystal polymers~\citep{wang_multi-functional_2022}, shape-memory polymers~\citep{ze_magnetic_2020}, magnetorheological elastomers (MREs)~\citep{Kim_Nature2018,wu_multifunctional_2020,kim_magnetic_2022}, \textit{etc}. In particular, MREs, consisting of a polymeric matrix embedded with magnetic particles, can change their elastic modulus and morph their shape under magnetic actuation~\citep{danas_experiments_2012, Kim_Nature2018}. When these materials are embedded with hard-ferromagnetic particles (hard-MREs), the interaction between the applied field and the remnant magnetization of the particles can impose significant torques in the material, leading to large deflections or rotations~\citep{Kim_Nature2018,wu_multifunctional_2020,kim_magnetic_2022}. The extreme, reversible deformation of structures made of hard-MREs combined with a fast and remote magnetic actuation have recently been leveraged to demonstrate novel functionalities in soft robotics~\citep{Kim_Nature2018,Hu_Nature2018, Gu_NatCommun2020}, biomedical engineering devices~\citep{Kim_SciRobot2019,zhang_voxelated_2021}, flexible electronics~\citep{yan_soft_2021,kim_magnetic_2022}, and metamaterials~\citep{chen_reprogrammable_2021}. The functionality of those devices can also be customized through the design of a heterogeneous magnetization profile of the MRE~\citep{Kim_Nature2018,Lum_PNAS2016,alapan_reprogrammable_2020}. 

Significant effort has been dedicated to the modeling of hard-MREs towards the rational design of magneto-active systems. In a pioneering study, \cite{Zhao_JMPS2019} proposed a continuum theory for hard-magnetic soft materials, assuming that the material was saturated (with a relative permeability close to 1), such that the applied magnetic field in the domain occupied by the material could be treated as a known, independent field variable. Then, an asymmetric stress tensor was derived from the magnetic body torque imposed in the material due to the misalignment between the magnetization and the applied field. This \textit{torque-based} model was validated by several test cases on thin-walled structures deformed under pure magnetic loading~\citep{Kim_Nature2018, Zhao_JMPS2019}, and successfully applied in practical designs~\citep{Kim_SciRobot2019,kuang_magnetic_2021}. In a follow-up study, \cite{garcia-gonzalez_magneto-visco-hyperelasticity_2019} extended this torque-based model to the case of hard-magnetic viscoelastic materials. Still using the torque-based continuum theory, \cite{zhang_micromechanics_2020} investigated the micromechanics of hard-MREs with a focus on particle rotation and torque transmission in a representative volume element. 

More recently, \cite{mukherjee_explicit_2021} have proposed a full-field, microstructurally guided modeling framework for hard-MREs. The constitutive description was derived with an emphasis on thermodynamic consistency and with parameters determined through homogenization. It is important to highlight that this comprehensive framework can consider the self-field generated by the MRE and the interaction between the microscopic magnetic particles. Unlike preceding models, the magnetic hysteresis of hard-MREs was also considered in the constitutive relations, which enables capturing the nonlinear behavior of hard-MREs under strong magnetic actuation, even above the coercivity of the particles and during the magnetization process. Using this model, the authors analyzed the deformation of an inhomogeneously magnetized cantilever beam in a case where the self-field cannot be neglected. More importantly. Their homogenization simulations demonstrated that the magnetization of a bulk pre-magnetized hard-MRE with a moderately soft matrix is independent of its stretching deformation. This observation is in contradiction with the previous continuum description for the magnetization of a deformed hard-MRE body proposed in~\cite{Zhao_JMPS2019}; a critical fact that we will discuss in detail later in this manuscript.
Inspired by the work of~\cite{mukherjee_explicit_2021}, \cite{garcia-gonzalez_microstructural_2021} proposed a microstructural model to study the effect of dipole-dipole interactions on the deformation of hard-MREs, in which only the rotation of the rigid particles is considered to derive the potentials corresponding to the mutual interactions of the particles and the interactions between the external field and the particles.
By developing a more comprehensive framework, \cite{rambausek_computational_2022} studied the interplay between the viscoelasticity of the polymeric matrix and the ferromagnetic hysteresis of the particles in hard-MREs.

Building upon the modeling work highlighted above, there have been several studies on the mechanics of magnetic slender structural elements toward enhancing the function of hard-MRE structures through geometry. In pioneering work on thin beams, \cite{Lum_PNAS2016} developed a 1D model to guide the design of heterogeneous magnetization profiles along the beam axis in order to realize a desired, complex deformation mode~\citep{Hu_Nature2018}. Starting from the torque-based continuum model, \cite{Wang_JMPS2020} presented a nonlinear theory of hard-magnetic elastica, which was then used in the design of 1D soft continuum robots~\citep{wang_evolutionary_2021}. The numerical implementation and experimental validation of this beam model under a gradient field were performed by \cite{yan_beams_2021} in a comprehensive study combining reduced-order modeling, 3D finite element modeling, and experiments. A detailed literature review of recent work on magnetic beams can be found in \cite{yan_beams_2021}. By studying 1D structural elements subjected to both bending and twisting, \cite{sano_rods_2021} derived a Kirchhoff-like theory to describe the 3D deformation of hard-magnetic rods. Using this model, the authors investigated the instability of straight and helical rods under magnetic actuation. In addition to beam-like elements, \cite{yan_magneto-active_2021} developed a 1D axisymmetric model for shells, which was used to rationalize the change of buckling strength of pressurized spherical shells when applying a magnetic field. This 1D shell model was later generalized in a 3D configuration by \cite{pezzulla_shells_2021}.

The aforementioned reduced-order models for beams, rods, and shells were all developed using the torque-based theory of hard-MREs proposed by~\cite{Zhao_JMPS2019}, ought to its simplicity and proven validity in several configurations~\citep{Kim_Nature2018,Kim_SciRobot2019,kim_magnetic_2022}. In this existing torque-based 3D theory, as mentioned above, the magnetization of the MRE body in the deformed configuration is related to that in the initial configuration through the deformation gradient, including both stretching and rotational contributions. However, the homogenization simulations performed recently by~\cite{mukherjee_explicit_2021} on a representative volume element (RVE) showed that the magnetization of hard-MREs remains constant under uniaxial tension perpendicular to pre-magnetization, since the polymeric matrix is much more compliant to the stretching deformation than the metallic particles (which are almost undeformed). The authors concluded that the magnetization of a hard-MRE depends only on its rotation, independently of stretch~\citep{mukherjee_explicit_2021,garcia-gonzalez_microstructural_2021}.
This microscopic composite effect was not taken into account by~\cite{Zhao_JMPS2019}. Hence, in some cases, the inaccurate description of magnetization might yield an erroneous magnetic part of the Cauchy stress, leading to unreliable predictions on the deformation of the MRE, as we demonstrate in the present study. This issue calls for an effort to update the existing torque-based theory~\citep{Zhao_JMPS2019} by incorporating the composite effect~\citep{mukherjee_explicit_2021}. Coming back to slender structures, plates are an essential member of the family in addition to beams, rods, and shells. Even if several promising applications involving magnetic plates have been proposed in the literature~\citep{kim_magnetic_2022,Hu_Nature2018,tang_versatile_2018,gray_review_2014}, their design has been primarily driven by intuition. Thus, there is a timely need for a plate theory to help understand the mechanics of plates under magnetic actuation and support the predictive and rational design of plate-like magneto-elastic structures.

Here, we propose a reduced-order theory for thin plates made of hard-MREs subjected to magnetic and mechanical loading. We start by modifying the existing continuum theory of~\cite{Zhao_JMPS2019} by taking into account the independence of magnetization on the stretching deformation of hard-MREs, as pointed out by~\cite{mukherjee_explicit_2021}. We derive a plate model based on this modified 3D theory by following a dimensional reduction procedure. Finally, our new 3D and plate theories are validated by comparing simulation results against precision experiments. We shall consider plates actuated by a uniform magnetic field and/or pressurized while systematically varying the loading parameters and boundary conditions. Both the modified 3D model and the developed plate model yield excellent predictions on the deformation of magnetic plates, establishing a theoretical and computational basis for the future modeling and design of slender plate-like structures. We also discuss the difference between our models and the existing theory, comparing their performances in the different test cases. We note that, to the best of our knowledge, this is the first time a rotation-based model has been validated experimentally for hard-MREs, here using a plate geometry.

Our paper is organized as follows. We define the problem in Section~\ref{sec:problem}. In Section~\ref{sec:3Dmodel}, for completeness, we review the existing torque-based 3D continuum theory~\citep{Zhao_JMPS2019} and then propose a modified description for the magnetization of a deformed hard-MRE body, yielding a rotation-based magnetic potential. In Section~\ref{sec:plate_model}, a plate model is then developed through dimensional reduction of the modified 3D theory. For comparison, we also provide a plate model derived from the existing 3D theory. The numerical implementation is described in Section~\ref{sec:numerics}, followed by our experimental methodology in Section~\ref{sec:experiments}. In Section~\ref{sec:results}, we present the experimental results and the comparison between the 3D and 2D models. Finally, the main findings are summarized and discussed in Section~\ref{sec:conclusion}.

\section{Definition of the problem}
\label{sec:problem}

\begin{figure}[b!]
\centering
 \includegraphics[width=0.7\columnwidth]{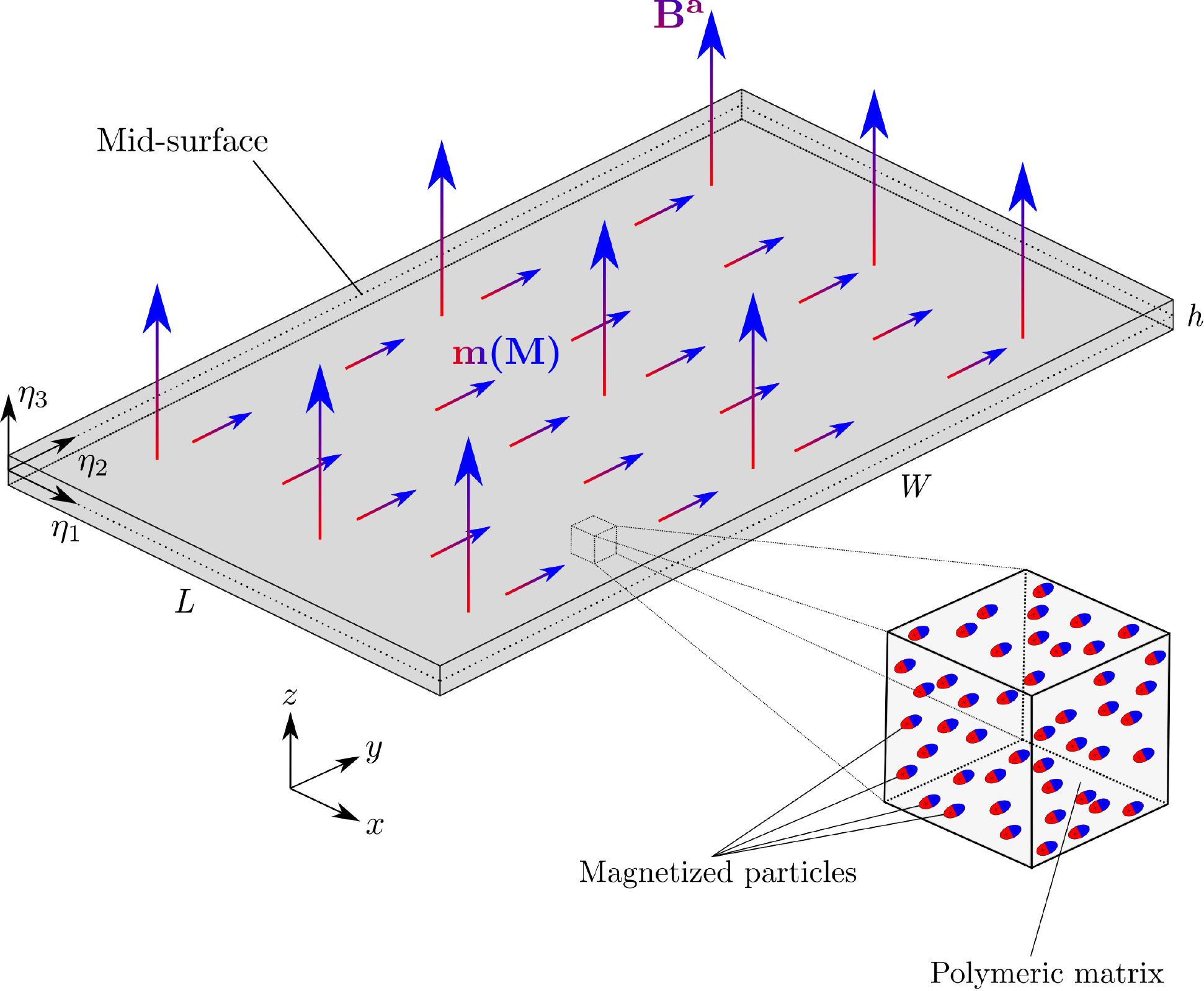}
 \caption{Schematic of a thin plate made of a hard-MRE in its initial flat configuration, subjected to an external magnetic field $\mathbf{B}^\mathrm{a}$. The magnified representative volume element (inset) illustrates the composition of the hard-MRE. The hard-magnetic particles are assumed to have been pre-magnetized, producing the remnant magnetization of the hard-MRE, $\mathbf{m}(\mathbf{M})$.}
 \label{fig:Fig1}
\end{figure}

In this section, we define the problem of the deformation of a hard-MRE plate under an applied magnetic field. In particular, we introduce the parameters of interest, clarify the modeling assumptions, and specify the research questions addressed in this work.

We consider a thin plate of length~$L$, width~$W$, and thickness~$h$ (with $L/h,W/h \gg 1$), as schematized in Fig.~\ref{fig:Fig1}. The absolute Cartesian coordinate system is denoted by $(\mathbf{{e}}_x, \mathbf{{e}}_y,\mathbf{{e}}_z)$. The plate is parameterized by the in-plane coordinates $\eta_1 \in [0,\,L]$ and $\eta_2\in [0,\,W]$ in the length and width directions, respectively, and by the out-of-plane coordinate $\eta_3 \in [-h/2,\,h/2]$ along the thickness. The initial flat configuration of the plate is represented by $\mathbf{X}=(\eta_1, \eta_2, \eta_3)$, and the deformed configuration by $\mathbf{x}(\eta_{1}, \eta_{2}, \eta_{3})$. Throughout the manuscript, capital letters denote quantities written with respect to the original configuration, and lowercase variables refer to the deformed configuration.

The plate is made of a hard-MRE with Young's modulus $E$ and Poisson's ratio $\nu$. The MRE is a two-phase composite made of micron-sized hard-magnetic particles embedded in a non-magnetic elastomeric matrix (see the inset of Fig.~\ref{fig:Fig1}). We assume that the diameters of the particles are much smaller than any dimensions of the plate, and that the particles are dispersed evenly in the matrix, such that the hard-MRE can be regarded as a homogeneous material. Throughout the present study, we shall only consider particles that have been pre-magnetized upon saturation with a permeability close to that of vacuum (relative permeability $\approx 1$). Consequently, in the initial configuration $\mathbf{X}$, given the magnetization of the individual particles, $\mathbf{M}^\mathrm{p}$, and their volume fraction, $c$, the effective magnetization of the MRE is
\begin{equation}
\mathbf{M}=c\mathbf{M}^\mathrm{p}\,.
\end{equation}
The counterpart of $\mathbf{M}$ in the deformed configuration $\mathbf{x}$ is denoted by $\mathbf{m}$. The magnetization of the MRE is assumed to be independent of external magnetic fields, provided that the MRE is operated under a field strength much smaller than the coercivity of the particles~\citep{Zhao_JMPS2019}. Furthermore, since we are considering thin plates, we can assume $\mathbf{M}$ is constant throughout the thickness. 

Under the application of a uniform magnetic field of flux density $\mathbf{B}^\mathrm{a}$, magnetic body torques $\boldsymbol{\tau}=\mathbf{m}\times\mathbf{B}^\mathrm{a}$ are exerted on the plate. These distributed torques arise from the interaction between the hard-magnetic particles and the external field, transmitted through the matrix. The plate deforms to the current configuration $\mathbf{x}$, with a tendency to align its magnetization to the applied field. We assume that the matrix is moderately soft~\citep{mukherjee_explicit_2021}, such that the rotations of the microscopic particles and the macroscopic MRE remain affine~\citep{vaganov_effect_2018,vaganov_training_2020,schumann_reversible_2021,zhang_micromechanics_2020}. 

In the following sections, we develop a plate model to describe the deformation of a hard-MRE plate under a uniform magnetic field. To do so, we shall first reformulate the magnetic potential of the existing torque-based 3D continuum theory by~\cite{Zhao_JMPS2019}, such that it can account for the composite effect recently studied by ~\cite{mukherjee_explicit_2021}. Specifically, we propose a model where the magnetization in the deformed configuration is related to the initial configuration by the rotation tensor, decomposed from the deformation gradient. We will then perform a dimensional reduction of the 3D theory to derive a 2D plate model written with respect to the mid-surface of the plate. Both the reformulated 3D theory and the developed plate model will be validated through precision experiments and compared to the existing 3D theory and the corresponding plate model for completeness.


\section{Continuum model of hard-MREs}
\label{sec:3Dmodel}
We start by proposing an improved 3D continuum model for hard-MREs, based on which a reduced-order plate model will then be established in Section~\ref{sec:plate_model}. First, for the sake of completeness, in Section~\ref{sec:3Dmodel_existing}, we review the torque-based model proposed by~\cite{Zhao_JMPS2019}. Then, in Section~\ref{sec:3Dmodel_modified}, we modify the magnetic potential of the latter model by taking into account the stretch-independence of magnetization reported by~\cite{mukherjee_explicit_2021}. 

\subsection{The existing torque-based model}
\label{sec:3Dmodel_existing}

In the pioneering work of~\cite{Zhao_JMPS2019}, the behavior of a bulk hard-MRE under an applied magnetic field was described by a Helmholtz free energy density, including an elastic part, ${U}^\mathrm{e}$, and a magnetic part, ${U}^\mathrm{m}$, as
\begin{equation}
U={U}^\mathrm{e}+{U}^\mathrm{m}\,,
\label{eq:U_total}
\end{equation}
written with respect to the initial configuration. Equilibrium requires that the variation of the total energy density $U$ integrated over the initial volume $V$ occupied by the hard-MRE body, $\mathcal{U}= \int_V U \mathrm{d}V $, must vanish for any kinematically admissible deformations; $\delta \mathcal{U} =0$. This yields the equilibrium equation and the corresponding weak formulation needed to solve for the deformation of the MRE body (analytically or numerically), under given boundary conditions. 

The magnetic part of the energy density, ${U}^\mathrm{m}$, introduced in Eq.~\eqref{eq:U_total} is defined as the potential of the magnetic body torque, $\boldsymbol{\tau}=\mathbf{m}\times\mathbf{B}^\mathrm{a}$, imposed on the MRE by the applied field, $\mathbf{B}^\mathrm{a}$, which, in the deformed configuration, is written as:
\begin{equation}
{u}^\mathrm{m}=-\mathbf{m}\cdot\mathbf{B}^\mathrm{a}\,,
\label{eq:Um_m}
\end{equation}
where $\mathbf{m}$ corresponds to the magnetization of the deformed hard-MRE body. \cite{Zhao_JMPS2019} proposed the relation between $\mathbf{m}$ and its counterpart in the initial configuration, $\mathbf{M}$, to be
\begin{equation}
\mathbf{m}=J^{-1}\mathbf{F}\mathbf{M}\,,
\label{eq:m_existing}
\end{equation}
where $\mathbf{F}$ is the deformation gradient and $J=\det(\mathbf{F})$ measures the relative volume change. Then, the density of magnetic potential with respect to the initial configuration is
\begin{equation}
{U}^\mathrm{m}_\mathbf{F}=-\mathbf{F}\mathbf{M}\cdot\mathbf{B}^\mathrm{a}\,.
\label{eq:Um_M_exsiting}
\end{equation}
This $\mathbf{F}$-based magnetic potential couples the elastic deformation and the magnetic actuation through the deformation gradient $\mathbf{F}$; for clarity, the subscript $\mathbf{F}$ is added in ${U}^\mathrm{m}_\mathbf{F}$, which is to be contrasted to ${U}^\mathrm{m}_\mathbf{R}$ introduced below in Eq.~\eqref{eq:Um_M_modified}.

The elastic energy density, ${U}^\mathrm{e}$, introduced in Eq.~\eqref{eq:U_total} describes the elastic response of the MRE. Consistently with~\cite{Zhao_JMPS2019}, in the 3D models throughout the present study, we use a neo-Hookean hyperelastic energy density
\begin{equation}
{U}^\mathrm{e}=\frac{G}{2}(J^{-{2}/{3}}I^\mathbf{C}_1-3)+\frac{K}{2}\left(J-1\right)^2\,,
\label{eq:Ue}
\end{equation}
where $G$ and $K$ are the shear and bulk moduli of the MRE, respectively, and $I^\mathbf{C}_1=\mathrm{tr}(\mathbf{C})$ is the first invariant of the right Cauchy–Green deformation tensor
\begin{equation}
\mathbf{C}=\mathbf{F}^\mathrm{T}\mathbf{F}\,.
\label{eq:C_3D}
\end{equation}

\subsection{A modified magnetic potential}
\label{sec:3Dmodel_modified}

We now examine the existing continuum model described above, but considering the fact that the MRE is a \textit{composite} material, with its effective magnetization arising from the dispersed magnetized particles. Equation~\eqref{eq:m_existing} states that the magnetization of a hard-MRE in its deformed configuration is determined by the deformation gradient, $\mathbf{F}$. From continuum mechanics, we know that the deformation gradient describes the local deformation due to both stretches and rotations, which are characterized by the stretch tensor $\mathbf{U}$ and  rotation tensor $\mathbf{R}$, respectively, and can be decomposed from $\mathbf{F}$ through polar decomposition as
\begin{equation}
\mathbf{F}=\mathbf{R}\mathbf{U}\,.
\label{eq:FRU_3D}
\end{equation}
Hence, Eq.~\eqref{eq:m_existing} indicates that both rotations and stretches should contribute to the magnetization $\mathbf{m}$ of the deformed hard-MRE body. However, as mentioned in Section~\ref{sec:introduction}, this continuum description is inconsistent with recent observations in homogenization simulations by~\cite{mukherjee_explicit_2021}. Specifically, in a microscopic RVE subjected to uniaxial tension, the stretching deformation occurs only in the matrix phase, since its modulus is orders of magnitudes lower than that of the particles. Consequently, the magnetization of the individual particles is independent of the stretch of the MRE, and the effective magnetization of the MRE should depend only on rotations~\citep{mukherjee_explicit_2021}.

Therefore, there is a need to modify the description of the magnetization $\mathbf{m}$ in Eq.~\eqref{eq:m_existing}, taking into account the fact that $\mathbf{m}$ is related only to the \textit{rotation} of the MRE. To do so, we replace the deformation gradient $\mathbf{F}$ in Eq.~\eqref{eq:m_existing} with the rotation tensor $\mathbf{R}$, yielding
\begin{equation}
\mathbf{m}=J^{-1}\mathbf{R}\mathbf{M}=J^{-1}(\mathbf{F}\mathbf{U}^{-1})\mathbf{M}\,,
\label{eq:m_modified}
\end{equation}
where we used the relation $\mathbf{R}=\mathbf{F}\mathbf{U}^{-1}$ from Eq.~\eqref{eq:FRU_3D}. The expression of the inverse of the stretch tensor $\mathbf{U}^{-1}$ as a function of $\mathbf{F}$ has been derived by~\cite{hoger_determination_1984}, \cite{sawyers_comments_1986} and \cite{guan-suo_determination_1998}, which, for convenience of the reader, are provided in \ref{Appendix_A}.
Finally, the magnetic potential corresponding to the magnetization in Eq.~\eqref{eq:m_modified} is 
\begin{equation}
{U}^\mathrm{m}_\mathbf{R}=-\mathbf{R}\mathbf{M}\cdot\mathbf{B}^\mathrm{a}\,.
\label{eq:Um_M_modified}
\end{equation}

Thus far, we have revised the magnetic potential proposed by~\cite{Zhao_JMPS2019} for hard-MREs by reformulating the magnetization $\mathbf{m}$ in the deformed configuration. The new expression of $\mathbf{m}$, which is independent of the stretching deformation of the MRE, fixes the inconsistency between the existing theory and the observation in homogenization simulations by~\cite{mukherjee_explicit_2021}. Hereafter, for simplicity, we shall refer to the continuum theory based on the existing magnetic potential in Eq.~\eqref{eq:Um_M_exsiting} as the ``\textit{$\mathbf{F}$-based 3D theory}", while the continuum theory based on the modified magnetic potential in Eq.~\eqref{eq:Um_M_modified} is referred to as the ``\textit{$\mathbf{R}$-based 3D theory}". The validity of the $\mathbf{R}$-based 3D theory and the significance of the modification proposed in this section will be demonstrated in Section~\ref{sec:results} through direct comparison against precision experiments on magnetic plates. 

\section{Reduced-order theory for hard-magnetic plates}
\label{sec:plate_model}

We proceed to develop a reduced-order model for thin plates made of hard-MREs, based on the $\mathbf{R}$-based 3D theory proposed in Section~\ref{sec:3Dmodel}. The general formulation for bulk hard-MRE materials is specialized to thin plates through dimensional reduction, yielding a 2D description. To do so, we first use the classic kinematics of thin plates in Section~\ref{sec:kinematics} to compute the rotation tensor of the mid-surface. Then, in Section~\ref{sec:plate_model_elastic_energy}-\ref{sec:plate_model_total_energy}, we reduce the magnetic and mechanical components of the energy of a plate from 3D to 2D with respect to the mid-surface. At last, in Section~\ref{sec:plate_model_F}, we provide another plate model reduced from the $\mathbf{F}$-based 3D theory, which is used in Section~\ref{sec:results} for comparisons with the model derived from the $\mathbf{R}$-based 3D theory.

\subsection{Rotation tensor of the mid-surface}
\label{sec:kinematics}

The kinematics of thin plates is well-established in classic plate theories, of which the F\"{o}ppl–von K\'{a}rm\'{a}n theory~\citep{Audoly_ElasticityGeometry2010} is arguably one of the most established. The focus of this section is to specialize the rotation tensor $\mathbf{R}$ for a thin plate from its deformation gradient $\mathbf{F}$, in order to compute the $\mathbf{R}$-based magnetic potential using Eq.~\eqref{eq:Um_M_modified}, noting that this is not a standard procedure in classic plate theory.
Taking the standard Kirchhoff-Love assumption~\citep{Audoly_ElasticityGeometry2010} that normals to the mid-surface remain normal and unstretched during deformation, we can write the deformed configuration $\mathbf{x}$ of the plate in terms of the mid-surface and its normal vector as
\begin{equation}
{\mathbf{x}}=\mathbf{r}+\eta_3\mathbf{n}\,,
\label{eq:deformed_config}
\end{equation}
where 
\begin{equation}
\mathbf{r}=(\eta_1+u_1, \eta_2+u_2,w)\,
\end{equation}
is the position vector of the mid-surface of the deformed plate, as a function of the in-plane displacements $u_1(\eta_1, \eta_2)$ and $u_2(\eta_1, \eta_2)$, and out-of-plane displacement $w(\eta_1, \eta_2)$ of the mid-surface. The vector normal to the deformed mid-surface is
\begin{equation}
{\mathbf{n}}=\frac{\mathbf{r}_{,1}\times\mathbf{r}_{,2}}{|\mathbf{r}_{,1}\times\mathbf{r}_{,2}|}\,,
\end{equation}
where~$(\cdot),_{\alpha}$ denotes partial derivation with respect to the (in-plane) mid-surface spatial coordinates $\eta_{\alpha}$ ($\alpha=1,2$). Hereon, Greek indices ($\alpha$, $\beta$, $\gamma$) shall represent the two in-plane directions.

Having defined the geometry of the initial and deformed configurations, we can compute the deformation gradient as
\begin{equation}
{\mathbf{F}}=\frac{\partial\mathbf{x}}{\partial\mathbf{X}}=\hat{\mathbf{F}}+\eta_3\mathbf{n}_{,\alpha}\otimes\mathbf{e}_{\alpha}\, ,
\label{eq:F_2D}
\end{equation}
where
\begin{equation}
\hat{\mathbf{F}}=\mathbf{r}_{,\alpha}\otimes\mathbf{e}_{\alpha}+\mathbf{n}\otimes\mathbf{e}_{3}
\label{eq:F_mid}
\end{equation}
is the deformation gradient of the mid-surface. Hereafter, hatted quantities $\hat{(\cdot)}$ are defined with respect to the mid-surface of the plate, thus independent of the out-of-plane coordinate $\eta_{3}$. The variation of ${\mathbf{F}}$ along the thickness of the plate is characterized by the term linear in $\eta_3$, in Eq.~\eqref{eq:F_2D}. The corresponding right Cauchy-Green deformation tensor of the mid-surface is
\begin{equation}
\hat{\mathbf{C}}=\hat{\mathbf{F}}^\mathrm{T}\hat{\mathbf{F}}=
\begin{pmatrix}
    \begin{matrix}\Bigg(\mathbf{r}_{,\alpha} \cdot \mathbf{r}_{,\beta}\Bigg)\end{matrix} & \begin{matrix} 0 \\ 0 \end{matrix} \\
    \begin{matrix} 0 & 0 \end{matrix} & \!\!1
\end{pmatrix}\,,
\label{eq:C_2D_mid}
\end{equation}
which measures the in-plane deformation of the mid-surface.

We now consider the rotation of the plate, which is required to describe the change of magnetization during deformation (see Eq.~\ref{eq:m_modified}). Note that we derive only the rotation tensor at the mid-surface, $\hat{\mathbf{R}}$, which determines the zero-order term in $\eta_{3}$ of the reduced magnetic potential (see Section~\ref{sec:plate_model_magnetic_potential_R}). 
According to Eq.~\eqref{eq:FRU_3D}, the stretch tensor $\hat{\mathbf{U}}$ at the mid-surface is needed in order to compute $\hat{\mathbf{R}}$. In Eq.~\eqref{eq:C_2D_mid}, we notice that off-diagonal components in the out-of-plane direction of $\hat{\mathbf{C}}$ vanish ($\hat{C}_{\alpha3} = \hat{C}_{3\alpha} = 0$), and that $\hat{C}_{33} = 1$, such that
\begin{equation}
\hat{\mathbf{U}}=\sqrt{\hat{\mathbf{C}}}=
\begin{pmatrix}
    \begin{matrix}\Bigg(\hat{U}_{\alpha\beta}\Bigg)\end{matrix} & \begin{matrix} 0 \\ 0 \end{matrix} \\
    \begin{matrix} 0 & 0 \end{matrix} & \!\!1
\end{pmatrix}\,.
\label{eq:U_2D_stretch}
\end{equation}
Eq.~\eqref{eq:U_2D_stretch} only requires the computation of the in-plane components~\citep{levinger_square_1980}:
\begin{equation}
\hat{U}_{\alpha\beta}=\sqrt{\hat{C}_{\alpha\beta}}=\left( \hat{C}_{\gamma\gamma} + 2\sqrt{\det(\hat{C}_{\alpha\beta})} \right)^{-\frac{1}{2}}
\left(\hat{C}_{\alpha\beta}+\sqrt{\det(\hat{C}_{\alpha\beta})}\delta_{\alpha\beta}\right)
\,.
\end{equation}
Finally, the rotation tensor of the mid-surface is given by
\begin{equation}
\hat{\mathbf{R}}=\hat{\mathbf{F}}\hat{\mathbf{U}}^{-1}\,,
\label{eq:R_2D}
\end{equation}
noting that it is a function of only $\eta_1$ and $\eta_2$, as expected for a reduced-order model.

Thus far, we have obtained the mid-surface quantities required to describe the deformation and rotation of a thin plate, including the deformation gradient $\mathbf{\hat{F}}$, right Cauchy-Green deformation tensor $\hat{\mathbf{C}}$, stretch tensor $\hat{\mathbf{U}}$, and rotation tensor $\hat{\mathbf{R}}$. These quantities will be used in the next two sections to derive the reduced elastic energy and the magnetic potential of the plate.

\subsection{Reduced elastic energy}
\label{sec:plate_model_elastic_energy}

Classic models for thin elastic plates typically involve the reduction of the 3D energy to a 2D description written with respect to the mid-surface of the plate. In the following, which can be found in the classic literature, we first consider the elastic energy, comprising a stretching term, $\hat{U}^\mathrm{s}$, and a bending term, $\hat{U}^\mathrm{b}$. For the sake of completeness, we provide the well-established expressions for $\hat{U}^\mathrm{s}$ and $\hat{U}^\mathrm{b}$ reduced from the 3D Kirchhoff-Saint Venant energy~\citep{Audoly_ElasticityGeometry2010},
\begin{equation}
\hat{U}^\mathrm{s}= \frac{Eh}{2(1-\nu^2)}\left[(1-\nu)E_{\alpha\beta} E_{\alpha\beta}+\nu E_{\alpha\alpha} E_{\beta\beta}\right]\,,
\label{eq:Us_2D}
\end{equation}
\begin{equation}
\hat{U}^\mathrm{b}= \frac{Eh^3}{24(1-\nu^2)}\left[(1-\nu)K_{\alpha\beta} K_{\alpha\beta}+\nu K_{\alpha\alpha} K_{\beta\beta}\right]\,,
\label{eq:Ub_2D}
\end{equation}
where $E$ and $\nu$ are the Young's modulus and Poisson's ratio of the material, respectively. The in-plane stretching strain in Eq.~\eqref{eq:Us_2D} is 
\begin{equation}
\hat{E}_{\alpha\beta} = \frac{1}{2}\left( \hat{C}_{\alpha\beta} - \delta_{\alpha\beta} \right) \,,
\end{equation}
and the bending strain in Eq.~\eqref{eq:Ub_2D} is
\begin{equation}
\hat{K}_{\alpha\beta} = -\mathbf{r}_{,\alpha} \cdot \mathbf{n}_{,\beta}\,.
\end{equation}
The reduced stretching and bending energy terms given above are valid for small strains and finite rotations, which are used in our numerical implementation of the plate model that we will present in Section~\ref{sec:numerics}.

\subsection{Reduced magnetic potential}
\label{sec:plate_model_magnetic_potential_R}

For the specialization of the 3D magnetic potential proposed in Eq.~\eqref{eq:Um_M_modified} to a thin plate, we need to substitute the rotation tensor $\mathbf{R}$ of Eq.~\eqref{eq:FRU_3D} into ${U}^\mathrm{m}_\mathbf{R}$ in Eq.~\eqref{eq:Um_M_modified}, and integrate over the thickness. Here, we take only the zero-order term of ${U}^\mathrm{m}_\mathbf{R}$ in $\eta_3$, \textit{i.e.}, the magnetic potential of the mid-surface. Therefore, the integral of ${U}^\mathrm{m}_\mathbf{R}$ over $\eta_3 \in [-h/2,\,h/2]$ yields the reduced magnetic potential of the plate:
\begin{equation}
\hat{U}^\mathrm{m}_\mathbf{R}=\int_{-h/2}^{h/2}{U}^\mathrm{m}_\mathbf{R}|_{\eta_3=0}\,\mathrm{d}\eta_3=-h(\hat{\mathbf{R}}\mathbf{M})\cdot\mathbf{B}^\mathrm{a}=-h[(\hat{\mathbf{F}}\hat{\mathbf{U}}^{-1})\mathbf{M}]\cdot\mathbf{B}^\mathrm{a}\,,
\label{eq:Um_2D_modified}
\end{equation}
which is linear to the thickness $h$. The adequacy of neglecting high-order terms in $\eta_3$ will be justified in Section~\ref{sec:results} through validation against experiments in a variety of test cases. Also, we remark that in recent work on hard-magnetic beams~\citep{yan_beams_2021}, rods~\citep{sano_rods_2021}, and shells~\citep{yan_magneto-active_2021}, the zero-order term of the magnetic potential has been shown to be sufficient in describing the mechanical behavior of these various structures under magnetic actuation.

\subsection{Plate model constructed using the \textbf{R}-based 3D theory}
\label{sec:plate_model_total_energy}

The summation of the 2D elastic and magnetic energy potentials integrated over the mid-surface $S$ yields the total energy of the plate
\begin{equation}
\hat{\mathcal{U}}_\mathbf{R}=\int_S \left(\hat{U}^\mathrm{s}+\hat{U}^\mathrm{b}+\hat{U}^\mathrm{m}_\mathbf{R}\right)\,\mathrm{d}S\,,
\label{eq:U_2D_exsiting}
\end{equation}
which is a functional of the displacement fields of the mid-surface ($u_1(\eta_1,\eta_2)$, $u_2(\eta_1,\eta_2)$, $w(\eta_1,\eta_2)$). Therefore, the problem has been reduced from a bulk description (3D) to the mid-surface (2D). Since the model is constructed from the $\mathbf{R}$-based 3D theory, we shall refer to it as the \textbf{R}-based plate model. Under given boundary conditions, we can solve for the displacements of the plate mid-surface by minimizing the energy functional in Eq.~\eqref{eq:U_2D_exsiting} , which, in this work, is conducted numerically using the finite element method (see Section~\ref{sec:numerics}).

\subsection{Alternative plate model derived from the \textbf{F}-based 3D theory}
\label{sec:plate_model_F}

Above, we obtained a model for thin hard-MRE plates reduced from the proposed \textbf{R}-based 3D theory introduced in Section~\ref{sec:3Dmodel}, where the magnetic potential depends on the rotation of the MRE. By following a procedure similar to that presented in Sections~\ref{sec:plate_model_elastic_energy}-\ref{sec:plate_model_total_energy}, a second alternative plate model can be derived from the existing \textbf{F}-based 3D theory~\citep{Zhao_JMPS2019}, in which both stretching and rotational deformations contribute to the magnetic potential. In the following, we provide this \textbf{F}-based plate model for the purpose of comparison, in the context of the case studies presented in Section~\ref{sec:results}, with the \textbf{R}-based plate model. This comparison will highlight an important drawback of the existing $\mathbf{F}$-based model.

Through a dimensional reduction on Eq.~\eqref{eq:Um_M_exsiting} of the \textbf{F}-based 3D theory (see \ref{Appendix_B} for details), we get the corresponding reduced magnetic potential written with respect to the mid-surface of the plate as
\begin{equation}
\hat{U}^\mathrm{m}_\mathbf{F}=-h(\hat{\mathbf{F}}\mathbf{M})\cdot\mathbf{B}^\mathrm{a}\,.
\label{eq:Um_2D_exsiting}
\end{equation}
Using the same stretching and bending energies as in Eqs.~\eqref{eq:Us_2D} and~\eqref{eq:Ub_2D}, respectively, the reduced total energy of the plate is
\begin{equation}
\hat{\mathcal{U}}_\mathbf{F}=\int_S \left(\hat{U}^\mathrm{s}+\hat{U}^\mathrm{b}+\hat{U}^\mathrm{m}_\mathbf{F}\right)\,\mathrm{d}S\,.
\end{equation}

Comparing $\hat{U}^\mathrm{m}_\mathbf{R}$ in Eq.~\eqref{eq:Um_2D_modified} with $\hat{U}^\mathrm{m}_\mathbf{F}$ in Eq.~\eqref{eq:Um_2D_exsiting}, we highlight  that $\hat{U}^\mathrm{m}_\mathbf{R}$ depends only on the rotation of the mid-surface $\hat{\mathbf{R}}$, independently of the stretch $\hat{\mathbf{U}}$. Hence, in general, $\hat{U}^\mathrm{m}_\mathbf{R} \neq \hat{U}^\mathrm{m}_\mathbf{F}$. We also notice that, for thin plates, $\hat{U}_{33}=1$ and $\hat{U}_{\alpha3} = \hat{U}_{3\alpha} = 0$ in Eq.~\eqref{eq:U_2D_stretch}, \textit{i.e.}, the normal and shear strains in the out-of-plane direction vanish under the Kirchhoff-Love assumption. Consequently, one can verify that, when the initial magnetization $\textbf{M}$ is parallel to $\mathbf{{e}}_z$, $\hat{U}^\mathrm{m}_\mathbf{R}$ and $\hat{U}^\mathrm{m}_\mathbf{F}$ are identical.
In Section~\ref{sec:results}, the two reduced-order plate models based on either $\hat{U}^\mathrm{m}_\mathbf{R}$ or $\hat{U}^\mathrm{m}_\mathbf{F}$ will be examined thoroughly and compared against experiments in a series of plate problems.

\section{Numerical implementation of the bulk and plate models}
\label{sec:numerics}

To validate the bulk (3D) and plate (2D) models proposed in Sections~\ref{sec:3Dmodel} and~\ref{sec:plate_model}, respectively, we will devise a set of physical and numerical experiments involving magnetic plates subjected to various loading conditions in Section~\ref{sec:results}. Before presenting the results, we introduce the numerical method we used to solve the plate problems. In summary, we write the total energy of the system, and solve for the displacement fields of the plate by minimizing the energy functional using the finite element method. The numerical implementation was performed using the commercial software package COMSOL Multiphysics. In the following, we provide details on the energy functional corresponding to each model and the parameters used in our numerical simulations. 

\subsection{Simulations using the $\mathbf{R}$-based and $\mathbf{F}$-based 3D models}

For the bulk (3D) simulations, we consider a 3D domain occupied by a plate of length $L$, width $W$, and thickness $h$. The total energy of the plate includes an elastic and a magnetic part. Specifically, for the elastic energy $U^\mathrm{e}$, we use the neo-Hookean model in Eq.~\eqref{eq:Ue}. For the magnetic potential, we choose either ${U}^\mathrm{m}_\mathbf{R}$ of Eq.~\eqref{eq:Um_M_modified} for the $\mathbf{R}$-based 3D theory proposed in Section~\ref{sec:3Dmodel_modified}, or ${U}^\mathrm{m}_\mathbf{F}$ of Eq.~\eqref{eq:Um_M_exsiting} for the $\mathbf{F}$-based 3D theory proposed by~\cite{Zhao_JMPS2019}.
In addition to elastic and magnetic effects, to capture the self-weight effects that are present in the experiments, we also take gravity into account through the potential 
\begin{equation}
{U}^\mathrm{g}=-\rho\mathbf{g}\cdot\mathbf{x}\,,
\label{eq:Ug_3D}
\end{equation}
where $\rho$ is the density of the material and $\mathbf{g}$ is the vector of gravitational acceleration. For cases when a uniform pressure is applied perpendicular to the surface of the plate (Sections~\ref{sec:example4} and~\ref{sec:example5}), we add the potential of a dead pressure $p$:
\begin{equation}
{U}^\mathrm{p}=-{p}{w}\,.
\label{eq:Up_3D}
\end{equation}
All in all, the energy functional considered in our numerics for the $\mathbf{R}$-based 3D theory is
\begin{equation}
\mathcal{U}_\mathbf{R}=\int_V \left({U}^\mathrm{e}+{U}^\mathrm{m}_\mathbf{R}+{U}^\mathrm{g}\right)\,\mathrm{d}V + \int_S {U}^\mathrm{p}\,\mathrm{d}S\,,
\label{eq:U_3D_FEM_R}
\end{equation}
and for the $\mathbf{F}$-based 3D theory, we have
\begin{equation}
\mathcal{U}_\mathbf{F}=\int_V \left({U}^\mathrm{e}+{U}^\mathrm{m}_{\mathbf{F}}+{U}^\mathrm{g}\right)\,\mathrm{d}V + \int_S {U}^\mathrm{p}\,\mathrm{d}S\,.
\label{eq:U_3D_FEM_F}
\end{equation}

We minimized $\mathcal{U}_\mathbf{R}$ or $\mathcal{U}_\mathbf{F}$ using the finite element method performed in COMSOL Multiphysics to solve for the unknown displacements of the plate ($u$, $v$, $w$), under given boundary conditions. To do so, the plate was discretized by a structured mesh with $40\times40\times8$ hexahedron elements seeded evenly in the length, width, and thickness directions, respectively. We used the cubic Hermite shape function. A mesh convergence study was conducted to ensure that the results were independent of the discretization. All geometric, elastic, and magnetic parameters used in the simulations were characterized independently in experiments (see Section~\ref{sec:experiments}), and, thus, there are no fitting parameters.

\subsection{Simulations using the $\mathbf{R}$-based and $\mathbf{F}$-based plate models}

In simulations using the dimensionally reduced $\mathbf{R}$-based and $\mathbf{F}$-based plate models, we consider only the 2D mid-surface of a plate. We provided the reduced stretching energy $\hat{U}^\mathrm{s}$ of the plate in Eq.~\eqref{eq:Us_2D} and the reduced bending energy $\hat{U}^\mathrm{b}$ in Eq.~\eqref{eq:Ub_2D}. The magnetic potential has been reduced to $\hat{U}^\mathrm{m}_\mathbf{R}$ in Eq.~\eqref{eq:Um_2D_modified} for the $\mathbf{R}$-based plate model, and to $\hat{U}^\mathrm{m}_\mathbf{F}$ in Eq.~\eqref{eq:Um_2D_exsiting} for the $\mathbf{F}$-based plate model. In addition, the 2D description of the gravitational potential of the plate is written as
\begin{equation}
\hat{U}^\mathrm{g}=-h\rho\mathbf{g}\cdot\mathbf{r}\,.
\label{eq:Ug_2D}
\end{equation}
When a pressure loading is imposed (Sections~\ref{sec:example4} and~\ref{sec:example5}), the potential of a dead pressure $p$ is identical to that given in Eq.~\eqref{eq:Up_3D}.

Collecting all of the above information, the energy functionals corresponding to the $\mathbf{R}$-based (Section~\ref{sec:plate_model_total_energy}) and $\mathbf{F}$-based (Section~\ref{sec:plate_model_F}) plate models are, respectively, 
\begin{equation}
\hat{\mathcal{U}}_\mathbf{R}=\int_S \left(\hat{U}^\mathrm{s}+\hat{U}^\mathrm{b}+\hat{U}^\mathrm{m}_\mathbf{R}+\hat{U}^\mathrm{g}+{U}^\mathrm{p}\right)\,\mathrm{d}S\,,
\label{eq:U_2D_FEM_R}
\end{equation}
and 
\begin{equation}
\hat{\mathcal{U}}_\mathbf{F}=\int_S \left(\hat{U}^\mathrm{s}+\hat{U}^\mathrm{b}+\hat{U}^\mathrm{m}_\mathbf{F}+\hat{U}^\mathrm{g}+{U}^\mathrm{p}\right)\,\mathrm{d}S\,.
\label{eq:U_2D_FEM_F}
\end{equation}
The corresponding simulations were performed using COMSOL Multiphysics by discretizing the 2D domain (plate mid-surface) with 6282 triangular elements (50 nodes on each edge) using the quintic Argyris shape function~\citep{argyris1968,pezzulla_shells_2021}, validated by a mesh convergence study. Constraints on displacements and their derivatives were specified on the edges of the mid-surface when required. The displacements predicted by the two plate models were obtained by minimizing  $\hat{\mathcal{U}}_\mathbf{R}$ and  $\hat{\mathcal{U}}_\mathbf{F}$ with respect to the displacement fields.

\section{Experimental fabrication, apparatus and protocols}
\label{sec:experiments}

Towards validating the theoretical models introduced in Section~\ref{sec:plate_model} and their numerical implementations presented in Section~\ref{sec:numerics}, we performed a series of precision experiments on thin plates made of hard-MREs, in a variety of test cases, and examined their deformations under different loading conditions. The detailed comparison between the numerical and experimental results will be provided in Section~\ref{sec:results}. In the present section, we elaborate on our experimental protocols for specimen fabrication, field generation, and deformation measurements. 

\subsection{Fabrication of hard-MRE plates}

\emph{Preparation of the MRE:} Following a well-established procedure~\citep{yan_beams_2021,yan_magneto-active_2021,Lum_PNAS2016}, our hard-MRE was fabricated by mixing vinylpolysiloxane (VPS) with NdPrFeB particles (average size $5\,\mu$m, MQFP-15-7-20065-089, Magnequench). Depending on the desired Young's modulus of the MRE, we used either VPS-22 (Elite Double 22, Zhermack, Young's modulus $\approx0.8\,$MPa) or VPS-32 (Elite Double 32, Zhermack, Young's modulus $\approx1.2\,$MPa). The \textit{NdPrFeB particles:VPS catalyst:VPS base} mass ratio was set to 2:1:1. Hereon, MREs made of VPS-22 or VPS-32 shall be referenced as MRE-22 or MRE-32, respectively.

\emph{Fabrication of the plate specimens:} The plate specimens were fabricated using the liquid MRE prepared as described above, through film application or casting. We used a thin film applicator (ZAA 2300, Zehntner) to fabricate the hard-MRE plates of thickness $481\le h \le 520\,\mu$m, as listed in Table~\ref{table:specimens}. We waited for $3\,$min after the mixture preparation to increase the viscosity to a desired value. The mixture was then poured onto a glass platform and spread by an automated moving applicator (speed $5\,$mm/s), with a gap height set to $750\,\mu$m. The MRE film was left to cure on the glass platform for $20\,$min. The final thickness depended on the combination of the mixture's viscosity, the applicator's moving speed, and the height of the gap between the applicator and the glass platform. The hard-MRE plate of thickness $h=869\,\mu$m (see Table~\ref{table:specimens}) was cast using a square acrylic frame placed on a glass platform as a mold. The liquid MRE was poured into the frame immediately after the preparation of the MRE mixture. The thickness was controlled by the volume of the liquid MRE poured.

\begin{table}[h!]
\small
\centering
\caption{Dimensions and properties of the specimens and levels of the magnetic field and air pressure applied}
\begin{adjustbox}{center}
\begin{tabular}{>{\centering\arraybackslash}p{1.7 cm}
>{\centering\arraybackslash}p{1.5cm}
>{\centering\arraybackslash}p{1.5cm} 
>{\centering\arraybackslash}p{1.5cm} 
>{\centering\arraybackslash}p{1.5cm} 
>{\centering\arraybackslash}p{1.5cm} 
>{\centering\arraybackslash}p{1.5cm}
>{\centering\arraybackslash}p{1.5cm}}
\cline{1-8}
{Section} & {$L$ {[}mm{]}} & {$h$ {[}$\mu$m{]}} & {$E$ {[}MPa{]}} & {$\rho$ {[}$\mathrm{g/cm^3}${]}} & {$M$ {[}kA/m{]}} & {$B^\textrm{a}$ {[}mT{]}} & {$p$ {[}Pa{]}}\\ \hline
{Section~\ref{sec:example1}} & 25.2 & 481 & 1.15 & 2.00 & 90.2 & $[0,20]$ & -\\ 
{Section~\ref{sec:example2}} & 42.0 & 869 & 1.81 & 2.03 & 91.3 & $[0,15]$ & -\\ 
{Section~\ref{sec:example3}} & 25.2 & 487 & 1.15 & 2.00 & 90.2 & $[0,60]$ & -\\ 
{Section~\ref{sec:example4}} & 25.0 & 520 & 1.07 & 2.00 & 90.2 & $[0,100]$ & $[0,300]$\\ 
{Section~\ref{sec:example5}} & 25.0 & 505 & 1.15 & 2.00 & 90.2 & $[0,160]$ & $[0,400]$ \\ \hline
\end{tabular}
\end{adjustbox}
\label{table:specimens}
\end{table}

Upon curing, the MRE plate was peeled off from the film applicator or the mold, and square specimens of the desired length $L$ were cut using a scalpel. The exact thickness of each specimen was measured using a microscope (VHX-5000, Keyence). In Table~\ref{table:specimens}, we report the dimensions of all the specimens used in the experiments together with the respective section where the corresponding results are presented and discussed. The specimens were then magnetized by saturating the embedded NdPrFeB particles up to 96\% under a strong axial magnetic field ($\approx2.5\,$T) generated by a pulse magnetizer (IM-K-010020-A, Magnet-Physik Dr. Steingroever). To do so, each specimen was positioned precisely in the magnetizing coil (MF-AsA-$\phi$120$\times$120,0mm, Magnet-Physik Dr. Steingroever) such that the desired orientation of magnetization (along or across the thickness of the plate, see Section~\ref{sec:results}) was aligned with the coil axis. 

\emph{Characterization of physical properties:} The volume fraction of the particles was set to $13.2\%$ for MRE-22 and $13.3\%$ for MRE-32, as computed from the mass ratio (1:1) and densities of the two phases using the law of mixtures~\citep{alger_polymer_2017} (VPS-22: $1.15\,\mathrm{g/cm^3}$ and VPS-32: $1.17\,\mathrm{g/cm^3}$, measured using a pycnometer, NdFeBrP: $7.61\,\mathrm{g/cm^3}$, provided by the supplier). The density of the MRE was computed to be $\rho=2.00\,\mathrm{g/cm^3}$ for MRE-22 and $\rho=2.03\,\mathrm{g/cm^3}$ for MRE-32, and the volume-average magnetization of the MRE was ${M} = 90.2\,$ kA/m for MRE-22 and ${M} = 91.3\,$kA/m for MRE-32, given the volume fraction and magnetization of the particles ($685.3\,$kA/m at 96\% saturation, as provided by the supplier).

The elastic properties of the MRE were characterized by mechanical testing of standard dogbone specimens under axial tension. Considering the evolution of Young's modulus over time, the dogbone tests and the plate experiments were both conducted at a specific time after the specimen fabrication. The Young's modulus of MRE-22 was measured to be $1.07\,$MPa at 48 hours, increasing to $1.15\,$MPa at 72 hours. For MRE-32, it was $1.81\,$MPa at 48 hours. The reported value is the average of 2 independent tests on 2 dogbone specimens. The Young's moduli corresponding to the plate specimens used in Section~\ref{sec:results} are also summarized in Table~\ref{table:specimens}.

\subsection{Generation of the magnetic and pressure loading}

Next, we elaborate the methods we used to generate the magnetic and pressure loading, which caused the deformation of the plates during the experiments.

\emph{Uniform magnetic field:} The magnetic field applied on the plate specimens was generated by two circular electromagnetic coils set in a Helmholtz configuration, powered by two $1.5\,$kW DC power supplies (EA-PSI 9200-25 T, EA-Elektro-Automatik). This magnetic field was axial and uniform in the central region between the coils. The setup we used was similar to that reported in~\cite{yan_beams_2021} but with coils of different sizes. Here, we chose a pair of bigger coils (inner diameter $190\,$mm, outer diameter $283\,$mm, height $54\,$mm) or a pair of small coils (inner diameter $86\,$mm, outer diameter $152\,$mm, height $33\,$mm) depending on the size of the specimen. We ensured that the specimen deformed within the uniform region of the field. The field strength was characterized using a Teslameter (FH 55, Magnet-Physik Dr. Steingroever). The ranges of the field strength for each experiment are listed in Table~\ref{table:specimens} .

\emph{Pressure generation and monitoring:} We applied a uniform pressure on the plate surface through an air chamber connected to a pneumatic loading system (see Sections~\ref{sec:example4} and~\ref{sec:example5}). The pressure differential was produced by injecting (exacting) air into (from) the chamber using a syringe pump (NE-1000, New Era Pump Systems Inc.) at the constant flow rate of $0.8\,$mL/min to increase (or decrease) the internal pressure. This pressure differential was monitored by a pressure sensor (785-HSCDRRN005NDAA5, Honeywell International Inc.) at an acquisition rate of $8\,$Hz.

\subsection{Measurement of plate deflections}
\label{sec:measurement}

The deflection of the plates under the imposed magnetic and pressure loading was measured using a camera (Sections~\ref{sec:example1}-~\ref{sec:example3}) or a position sensor (Sections~\ref{sec:example4} and~\ref{sec:example5}), depending on its extent. For deflections much larger than the plate thickness, the deformed configuration was captured by a camera (Nikon D850). The displacements at different loading levels were then acquired through digital image processing in MATLAB. For deflections of the order of the plate thickness, an optical position sensor (CL6-MG20, CCS-Optima+, STIL) with a resolution of $0.8\,\mu$m was used to directly measure the out-of-plane displacement.

\section{Results: Validation of the $\mathbf{R}$-based and $\mathbf{F}$-based models against experiments}
\label{sec:results}

Having introduced all of our theoretical, numerical, and experimental machinery, we proceed by presenting a series of test cases designed to validate our models through precision experiments. Our goal is to comparatively evaluate the performance of the $\mathbf{R}$-based and $\mathbf{F}$-based models in predicting the deformation of hard-MRE plates in different loading conditions. We also aim to identify the conditions where the $\mathbf{R}$-based models outperform the $\mathbf{F}$-based models in terms of the accuracy of prediction. 

We have devised experiments in which thin plates are magnetized and loaded in various configurations, involving both rotations and stretches of the mid-surface. Specifically, in Sections~\ref{sec:example1} and~\ref{sec:example3}, we consider plates deformed under pure magnetic actuation. Then, in Sections~\ref{sec:example4} and~\ref{sec:example5}, we examine the response of a plate to pressure loading in the presence of a magnetic field. In each test case, the predictions from the 3D and 2D models developed in Sections~\ref{sec:3Dmodel} and~\ref{sec:plate_model} are compared against experimental measurements. We highlight that the geometric, elastic, and magnetic properties used in the simulations have been either characterized experimentally (see Section~\ref{sec:experiments}) or stated by providers, and, thus, there are no fitting parameters.

\subsection{Deflection of a suspended plate clamped at one edge}
\label{sec:example1}

\begin{figure}[t!]
\centering
 \includegraphics[width=0.75\columnwidth]{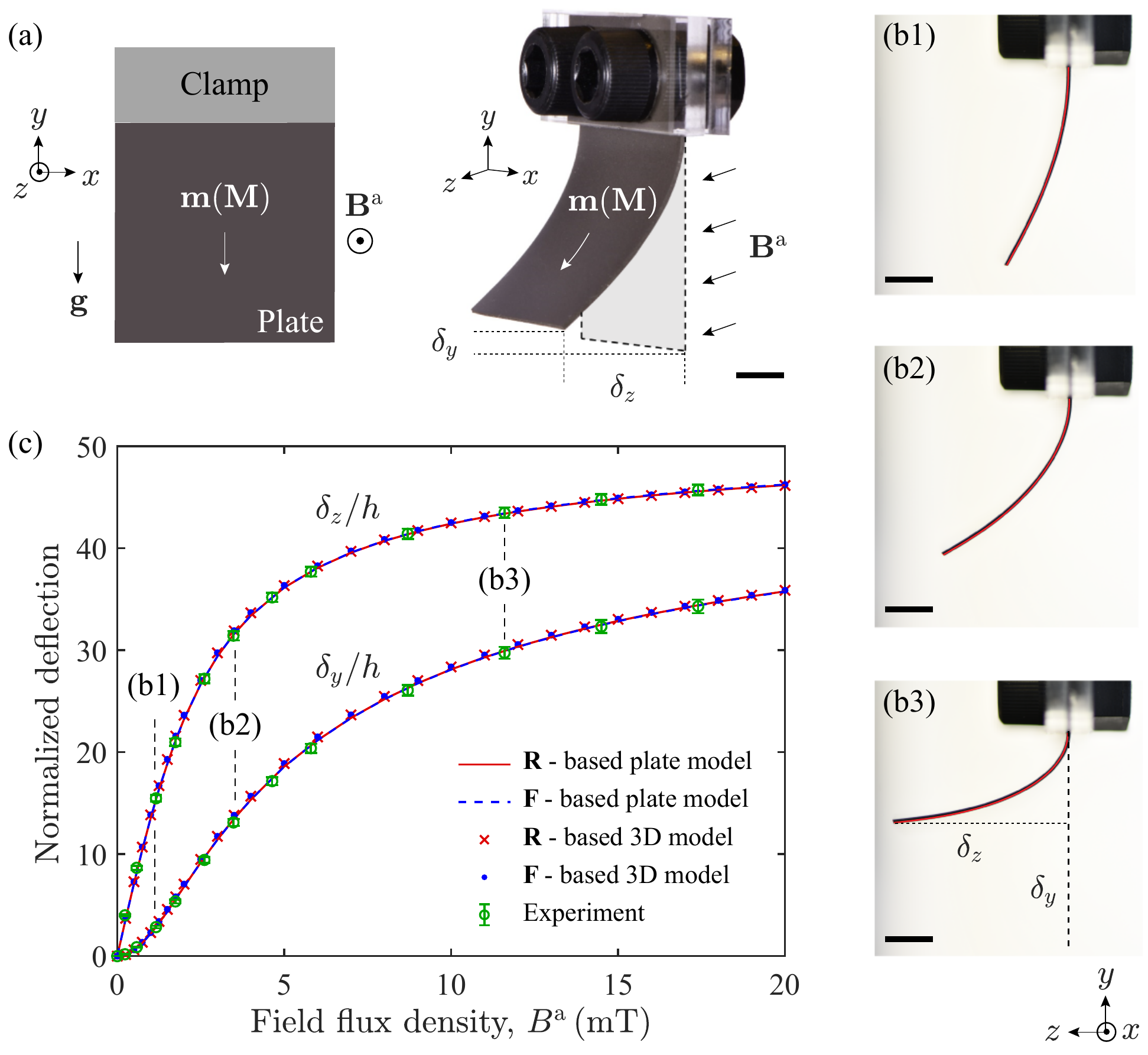}
 \caption{Deformation of a suspended square plate under a uniform magnetic field. (a) Left: Schematic diagram in the $x-y$ plane. Right: Photograph of a plate specimen tested, with its top edge clamped between rigid plates. The gravitational field $\mathbf{g}$ points in the $-y$ direction. (b) Experimental photographs (side view: $y-z$ plane) of the deformed plate at different levels of magnetic flux density: (b1) $B^\mathrm{a}=1.2\,$mT, (b2) $B^\mathrm{a}=3.5\,$mT, and (b3) $B^\mathrm{a}=12\,$mT. The red lines in each photograph are the deformed configurations predicted by the $\mathbf{R}$-based plate model. (c) Normalized components of the deflection at the bottom edge of the plate, $\delta_y/h$ and $\delta_z/h$, versus magnetic flux density $B^\mathrm{a}$. The error bars of the experimental data correspond to the standard deviations of the measurements on three identical specimens. The labeled points correspond to the photographs in (b1), (b2), and (b3) for the respective values of $B^\mathrm{a}$. The dimensions of the specimen are provided in Table~\ref{table:specimens}.}
 \label{fig:Fig2}
\end{figure}

First, we consider a square plate magnetized in-plane with $\mathbf{M} = -M\mathbf{{e}_{2}}$ and suspended under gravity, as shown in Fig.~\ref{fig:Fig2}(a). By applying a uniform magnetic field $\mathbf{B}^\mathrm{a} = \pm B^\mathrm{a} \mathbf{{e}_{3}}$, perpendicular to the plate magnetization in the initial configuration, magnetic torques are generated throughout the plate. Consequently, the plate deflects, as shown in the representative photographs of Fig.~\ref{fig:Fig2}(b1)-(b3), captured during the experiment at different levels of field strength. We performed the same test on three identical specimens, and each was tested twice with the field in the $z$ or $-z$ directions, respectively, to reduce the error from positioning the plate. The absolute values of the deflection were extracted from the experimental photographs, and the reported data is the average of the six independent measurements.

In Fig.~\ref{fig:Fig2}(c), we plot the deflection of the bottom edge of the plate in the $y$ and $z$ directions, $\delta_y/h$ and $\delta_z/h$ (normalized by the plate thickness), as a function of the magnetic flux density $B^\mathrm{a}$. The curves are monotonic but highly nonlinear for large deflections, due to the geometric nonlinearity and the dependence of magnetic torques on the rotation of the plate. When the magnetization of the majority of the plate is aligned with the external field, the curves reach a plateau. In this first test case, both the $\mathbf{F}$-based and the $\mathbf{R}$-based models yield results in excellent agreement with the experiments. Since both the 3D and 2D models are geometrically nonlinear, the large deflections considered can be predicted accurately. In particular, we superimpose the deformed configurations as predicted by the $\mathbf{R}$-based plate model on the corresponding plate specimen in each photograph shown in Fig.~\ref{fig:Fig2}(b1)-(b3), observing excellent agreement throughout.  

\subsection{Deflection of a plate clamped at two edges}
\label{sec:example2}

\begin{figure}[b!]
\centering
 \includegraphics[width=0.75\columnwidth]{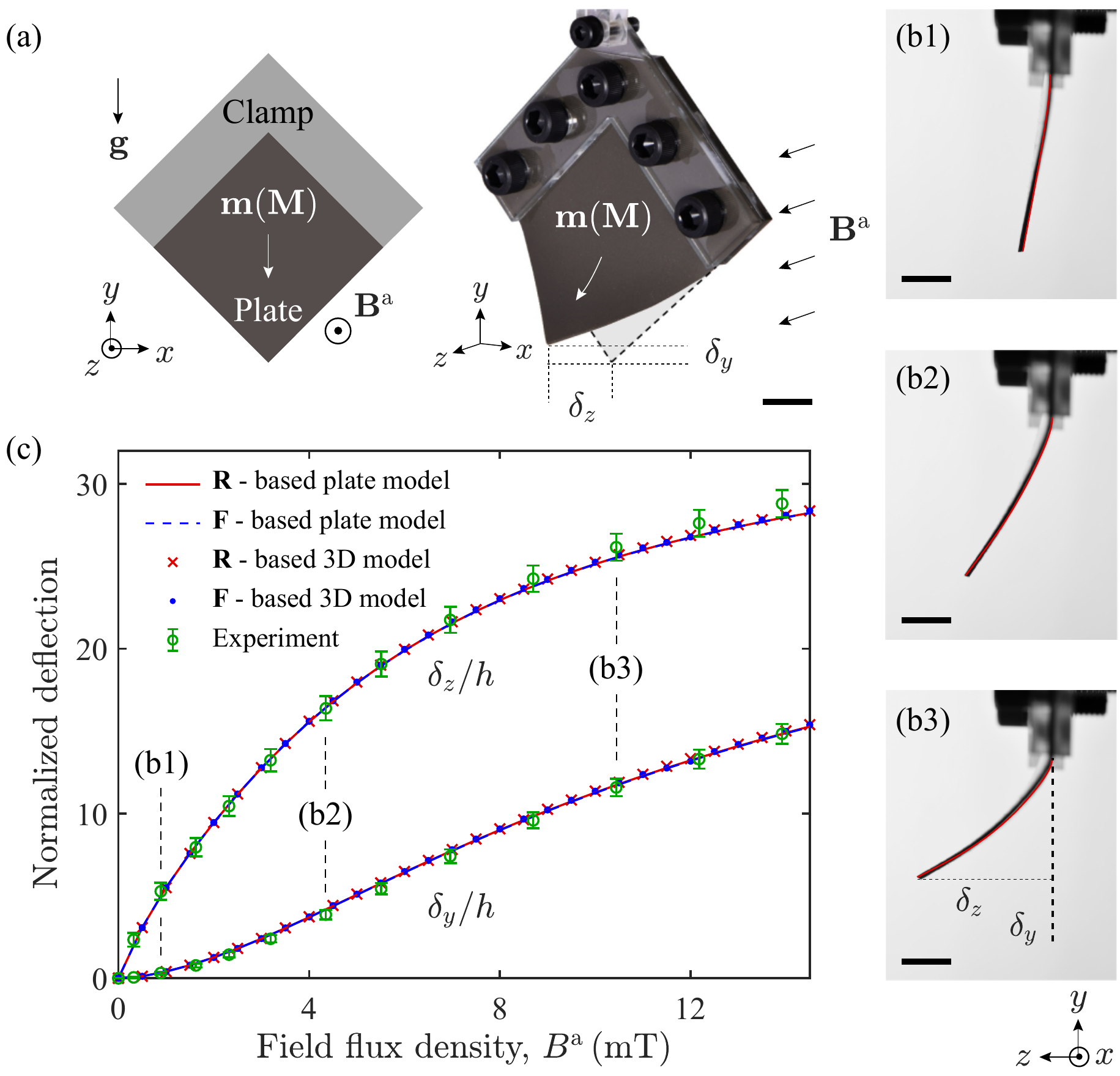}
 \caption{Deformation of a plate clamped at two edges under a uniform magnetic field. (a) Left: Schematic diagram in the $x-y$ plane. Right: Photograph of a plate specimen, with two edges clamped between rigid plates. The gravitational field $\mathbf{g}$ points in the $-y$ direction. (b) Experimental photographs (side view: $y-z$ plane) of the deformed plate at different levels of magnetic flux density: (b1) $B^\mathrm{a}=0.9\,$mT, (b2) $B^\mathrm{a}=4.4\,$mT, and (b3) $B^\mathrm{a}=10\,$mT. The red lines in each photograph are the deformed configurations predicted by the $\mathbf{R}$-based plate model. (c) Normalized components of the deflection at the free corner of the plate, $\delta_y/h$ and $\delta_z/h$, versus magnetic flux density $B^\mathrm{a}$. The error bars correspond to the standard deviations of the measurements on three identical specimens. The labeled points correspond to the photographs in (b1), (b2), and (b3) for the respective values of $B^\mathrm{a}$. The dimensions of the specimen are provided in Table~\ref{table:specimens}.}
 \label{fig:Fig3}
\end{figure}

In a second test case, we continue to probe our models for large deflections by considering a square plate clamped at two of its boundaries, as shown in Fig.~\ref{fig:Fig3}(a). The plate is pre-magnetized in-plane and diagonally, \textit{i.e.}, $\mathbf{M} = -M\mathbf{{e}_{2}}$. The lower part of plate deflects due to magnetic body torques imposed by a uniform magnetic field applied $\mathbf{B^{a}}$ in the out-of-plane direction. Figure~\ref{fig:Fig3}(b1)-(b3) show representative experimental photographs with increasing field flux density. We tested three identical specimens, each deflected in the $z$ and $-z$ directions under the field of $\mathbf{B^{a}} = B^{a}\mathbf{{e}_{3}}$ and $\mathbf{B^{a}} = -B^{a}\mathbf{{e}_{3}}$, respectively. The absolute values of the deflection were extracted from the experimental photographs. 

In Fig.~\ref{fig:Fig3}(c), we plot the average deflection at the free corner of the plate, $\delta_y/h$ and $\delta_z/h$ (normalized by its thickness in the $y$ and $z$ directions), as a function of the magnetic flux density $B^\mathrm{a}$. The deflections measured in this test case reach up to nearly 30 times the plate thickness, and are predicted well by all the models. This can also be seen in the comparison between the deformed configurations of the plate captured in the experiment and predicted by the $\mathbf{R}$-based plate model, as presented in Fig.~\ref{fig:Fig3}(b1)-(b3). 

Through this test case and the one in Section~\ref{sec:example1}, we have validated the ability of all our models to describe the large deflection of hard-MRE plates with geometric nonlinearities. In particular, we have demonstrated that our dimensional reduction procedure yields results just as accurate as the 3D models. Since the deformation considered so far is bending-dominated with negligible stretching deformation of the mid-surface, there is no significant difference between the $\mathbf{R}$-based and $\mathbf{F}$-based models.

\subsection{Deflection of a plate clamped at three edges}
\label{sec:example3}

\begin{figure}[t!]
\centering
 \includegraphics[width=0.75\columnwidth]{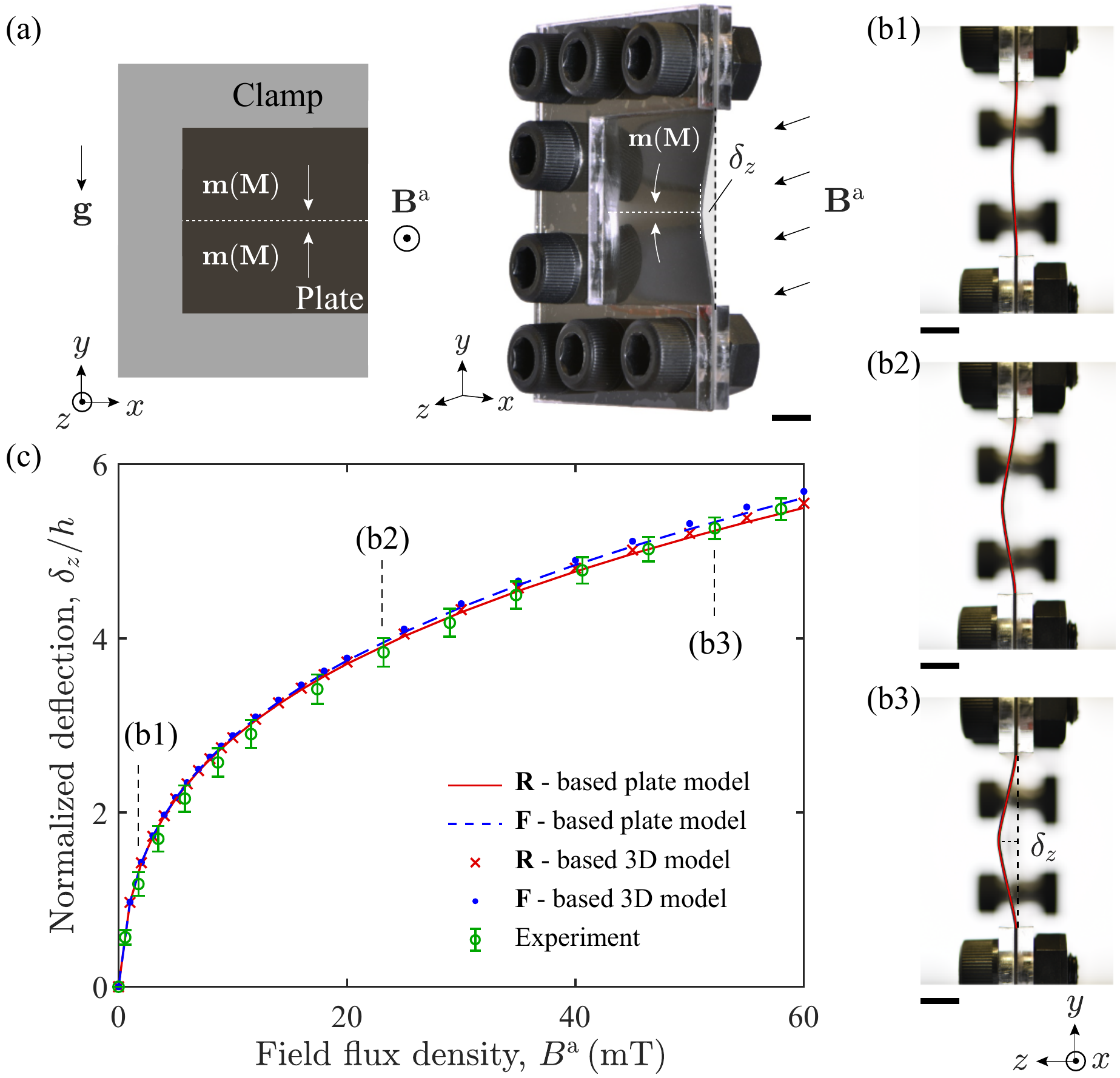}
 \caption{Deformation of a plate clamped at three edges under a uniform magnetic field. (a) Left: Schematic diagram in the $x-y$ plane. Right: Photograph of a plate specimen, with three edges clamped between rigid plates. The gravitational field $\mathbf{g}$ points in the $-y$ direction. (b) Experimental photographs (side view: $y-z$ plane) of the deformed plate at different levels of magnetic flux density: (b1) $B^\mathrm{a}=1.7\,$mT, (b2) $B^\mathrm{a}=23\,$mT, and (b3) $B^\mathrm{a}=52\,$mT. The red lines in each photograph are the deformed configurations predicted by the $\mathbf{R}$-based plate model. (c) Normalized deflection at the midpoint of the free edge, $\delta_z/h$, versus magnetic flux density $B^\mathrm{a}$. The error bars correspond to the standard deviations of the measurements on three identical specimens. The labeled points correspond to the photographs in (b1), (b2), and (b3) for the respective values of $B^\mathrm{a}$. The dimensions of the specimen are provided in Table~\ref{table:specimens}.}
 \label{fig:Fig4}
\end{figure}

From this test case onward, to further explore the potential of our $\mathbf{R}$-based models, the excited modes of deformation involve stretching of the plate mid-surface. Here, a square plate is clamped at three edges, as shown in Fig.~\ref{fig:Fig4}(a). We have introduced in-plane, symmetric magnetization vectors of equal magnitude facing the middle of the plate. Experimentally, this was achieved by folding the plate in half during magnetization. A uniform magnetic field $\mathbf{B^{a}}$ is applied in the out-of-plane direction to deflect the plate, inducing both bending and stretching deformations. We present representative experimental photographs in Fig.~\ref{fig:Fig4}(b1)-(b3) at three loading levels. Three identical specimens were tested under both $\mathbf{B^{a}} = B^{a}\mathbf{{e}_{3}}$ and $\mathbf{B^{a}} = -B^{a}\mathbf{{e}_{3}}$, and the deflections were extracted from experimental photographs. 

Figure~\ref{fig:Fig4}(c) presents the averaged deflection at the midpoint of the free edge, $\delta_{z}/h$ (normalized by the plate thickness), as a function of the magnetic flux density $B^\mathrm{a}$. Despite a good agreement with experiments, we notice a small difference between the $\mathbf{F}$-based and $\mathbf{R}$-based models, which becomes more prominent as the deflection increases. This discrepancy is due to the different descriptions on the magnetization of the deformed plate (see Eqs.~\ref{eq:m_existing} and~\ref{eq:m_modified}), as we have detailed in Section~\ref{sec:3Dmodel}. However, the differences in the predictions of the two families of models are not yet significant enough to state that the $\mathbf{R}$-based models perform better overall.

\subsection{Deflection of a pressurized, fully-clamped plate with in-plane magnetization}
\label{sec:example4}

\begin{figure}[t!]
\centering
 \includegraphics[width=1\columnwidth,center]{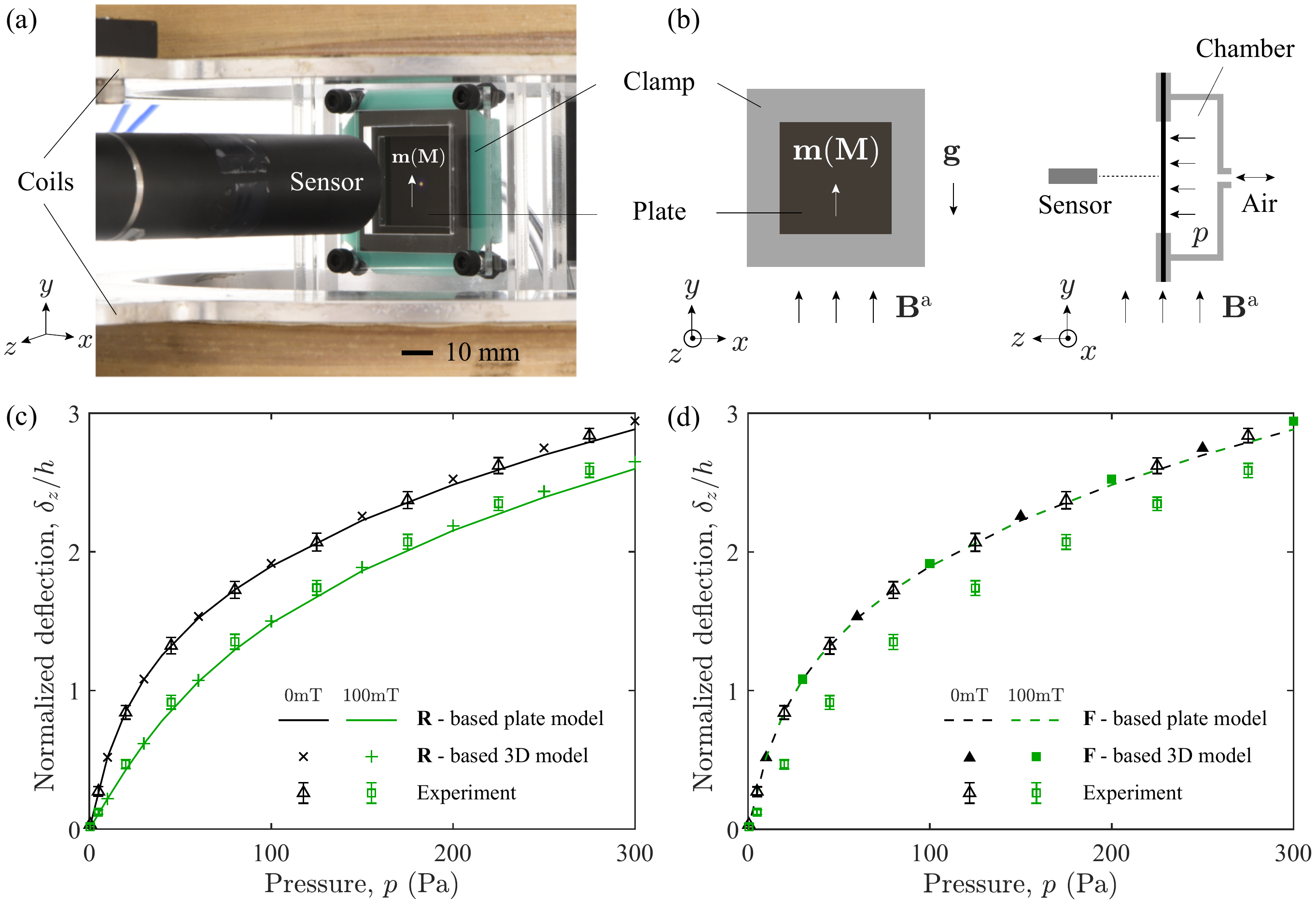}
 \caption{Deformation of a fully-clamped plate with \textit{in-plane} magnetization under combined pressure and magnetic loading. (a) Photograph of a plate specimen positioned in between the Helmholtz coils. The deflection at the plate center was measured by an optical position sensor. (b) Schematic of the configuration considered in the experiment. The pressure loading was imposed on the plate by changing the air pressure in the chamber. (c) Normalized deflection at the center of the plate, $\delta_z/h$, versus imposed pressure, $p$, at magnetic flux density $B^\mathrm{a}=\{0, 100, 200\}\,$mT. The error bars correspond to the standard deviations of the measurements on three identical specimens.}
 \label{fig:Fig5}
\end{figure}

We further the validation of our models by considering a test case where the external magnetic field is aligned with the plate magnetization in the initial configuration, which shall emphasize the difference between the $\mathbf{F}$-based and the $\mathbf{R}$-based models. To do so, we fully clamp a square plate on its four boundaries, set the magnetization $\mathbf{M} = M\mathbf{{e}_{2}}$, and apply a uniform field $\mathbf{B^\mathrm{a}} =  B^\mathrm{a}\mathbf{{e}_{2}}$ aligned with $\mathbf{M}$. Figure~\ref{fig:Fig5}(a) and~(b) show a photograph and a schematic of our experimental setup. The two vectors $\mathbf{B}^\mathrm{a}$ and $\mathbf{M}$ are oriented in-plane and parallel in the initial configuration. We then impose a uniform pressure on the plate to introduce both bending and stretching deformation in the presence of the magnetic field. The field strength is kept constant, but the pressure is swept from 0 to 300 Pa. The plate deflection induces an angle between $\mathbf{B}^\mathrm{a}$ and $\mathbf{M}$, resulting in magnetic torques that oppose the deformation. We quantify this pressure-induced, magnetically-restricted deformation by examining the deflection of the pressurized plate at different levels of magnetic flux density $B^\mathrm{a}$. Since, in this case, the deflection is smaller than that in the previous test cases, of the order of the plate thickness, we used a position sensor for maximal precision (see Section.~\ref{sec:measurement}).

Figure~\ref{fig:Fig5}(c) and~(d) show the deflection at the center of the plate normalized by the plate thickness, $\delta_z/h$, as a function of pressure, $p$, with and without the external magnetic field, \textit{i.e.}, $B^\mathrm{a}=\{0, 100\}\,$mT. We first focus on the experimental data and the predictions obtained through the $\mathbf{R}$-based 2D and 3D models in Fig.~\ref{fig:Fig5}(c). In this case, we observe excellent agreement between theory and experiments. We highlight the fact that at a given level of pressure loading, the plate deflection decreases with increasing magnetic flux density due to magnetic body torques acting against the rotation of the plate. 

Having validated the $\mathbf{R}$-based models, we turn to compare the experimental data with the predictions from the $\mathbf{F}$-based models. We can see from Fig.~\ref{fig:Fig5}(d) that, in the presence of a magnetic field (${B}^\mathrm{a}=100\,$mT), both the $\mathbf{F}$-based 3D and plate models deviate significantly from the experiments, even for small deflections ($\delta_z/h<1$). Specifically, the predicted deflections at ${B}^\mathrm{a}=100\,$mT show negligible changes from the case of ${B}^\mathrm{a}=0\,$mT. This disagreement suggests that the accuracy of the $\mathbf{F}$-based models worsens considerably when the plate deflects with stretching deformation in the presence of an applied field parallel to the magnetization. Our experiment on a hard-MRE plate constitutes a first step towards validating the independence of magnetization on stretching deformation reported by~\cite{mukherjee_explicit_2021} through homogenization simulations. The test case investigated in the next section will focus on further discriminating the two families of models.

\subsection{Deflection of a pressurized, fully-clamped plate with out-of-plane magnetization}
\label{sec:example5}

\begin{figure}[t!]
\centering
 \includegraphics[width=1\columnwidth,center]{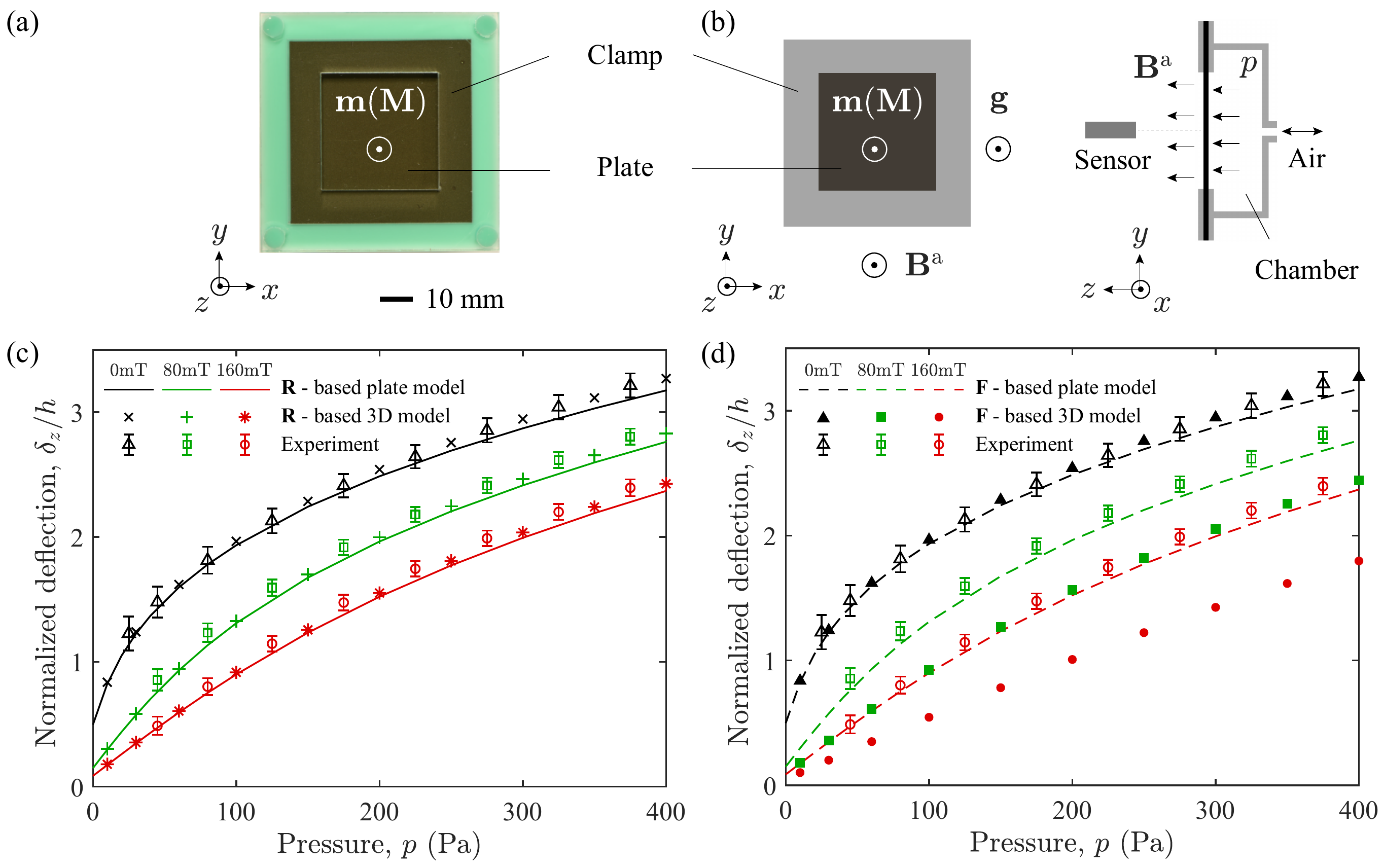}
 \caption{Deformation of a fully-clamped plate with \textit{out-of-plane} magnetization under combined pressure and magnetic loading. (a) Photograph of a plate specimen clamped at its four edges. (b) Schematic of the configuration considered in the experiment. The pressure loading was imposed on the plate by changing the air pressure in the chamber. (c) Normalized deflection $\delta_z/h$ at the center of the plate, versus imposed pressure, $p$, at different levels of applied magnetic flux density $B^\mathrm{a}=\{0, 80, 160\}\,$mT. The initial deflections in experiments at $p=0$ due to gravity were characterized using an optical profilometer (VR-3200, Keyence). The error bars correspond to the standard deviations of the measurements on three identical specimens.}
 \label{fig:Fig6}
\end{figure}

In the last test case investigated, we further examine the effect of stretching deformation when a magnetic field $\mathbf{B^\mathrm{a}}$ is applied parallel to the plate magnetization $\mathbf{M}$, but we now set $\mathbf{B^\mathrm{a}}$ and $\mathbf{M}$ in the out-of-plane direction. Figure~\ref{fig:Fig6}(a) and~(b) present a photograph and a schematic of the pressurized, fully-clamped plate with magnetization $\mathbf{M} = M \mathbf{{e}_{3}}$ under a magnetic field $\mathbf{B^\mathrm{a}} = B^\mathrm{a}\mathbf{{e}_{3}}$. Both $\mathbf{M}$ and $\mathbf{B^\mathrm{a}}$ are oriented perpendicularly to the plate mid-surface, as opposed to the configuration considered in Section~\ref{sec:example4}. 

Figure~\ref{fig:Fig6}(c) and~(d) present the normalized deflection at the center of the plate, $\delta_{z}/h$, versus the applied pressure, at three different fixed levels of magnetic flux density $B^\mathrm{a}=\{0, 80, 160\}\,$mT. The experimental results are accurately reproduced by the $\mathbf{R}$-based plate and 3D models (see Fig.~\ref{fig:Fig6}c). The magnetic torques induced due to the deflection increase the resistance of the plate to the pressure loading. This effect becomes more prominent for stronger applied fields.

By contrast, we find that, in Fig.~\ref{fig:Fig6}(d), the $\mathbf{F}$-based 3D model yields predictions that disagree with experimental data. In particular, the $\mathbf{F}$-based 3D model predicts a change in deflection with respect to the case of $B^\mathrm{a}=0\,$mT twice as large as that obtained in the experiment, and that given by the $\mathbf{R}$-based 3D and plate models. As reported by~\cite{Zhao_JMPS2019}, according to the $\mathbf{F}$-based theory, pre-stretches are induced in a free bulk MRE when the applied field is parallel to the magnetization. In our case, in-plane pre-stresses are induced in the mid-surface by the out-of-plane pre-stretch, since the plate is fully constrained. These magnetically-induced in-plane stresses, in addition to the magnetic torques due to the rotational deformation, further stiffen the plate, explaining the observation in Fig.~\ref{fig:Fig6}(d) that the $\mathbf{F}$-based 3D model underestimates the plate deflection. In contrast, the magnetically-induced pre-stretches and pre-stresses do not exist for the $\mathbf{R}$-based models, since the magnetic potential is independent of stretching deformation. Based on this comparison against experiments, we conclude that the pre-stretches and pre-stresses predicted by the $\mathbf{F}$-based theory are erroneous, and the source of the discrepancy with experiments. 

In Fig.~\ref{fig:Fig6}(d), we remark the $\mathbf{F}$-based plate model agrees unexpectedly well with the experiment, even though it is reduced from a 3D model that itself yields incorrect predictions. We attribute this contradiction to the Kirchhoff-Love assumption applied when deriving the kinematics of thin plates through dimensional reduction. This assumption eliminates any out-of-plane deformation, thus removing the artificial pre-stretch in the out-of-plane direction. Hence, no in-plane pre-stresses are induced in the plate mid-surface. Consequently, the $\mathbf{F}$-based plate model yields predictions in surprising agreement with the experimental data due to an artifact imposed by the Kirchhoff-Love assumption during the dimensional reduction.

\section{Conclusion}
\label{sec:conclusion}

We studied plates made of hard-MREs, and developed a reduced-order model to describe their mechanical deformation under combined magnetic and mechanical loading. Our plate model was derived by performing the dimensional reduction of a 3D continuum theory, adapted from the torque-based theory of hard-MREs proposed in the pioneering study of \cite{Zhao_JMPS2019}. The modified 3D theory takes into account the composite nature of the MRE. In particular, we related the magnetization vectors in the initial and deformed configurations using only the rotation tensor $\mathbf{R}$, decomposed from the deformation gradient $\mathbf{F}$, yielding a $\mathbf{R}$-based expression of the magnetic potential. Our reformulation was inspired by the recent homogenization simulation results of~\cite{mukherjee_explicit_2021}, which demonstrated that the magnetization of saturated hard-MREs is independent of stretching deformation.

We devised a series of test cases to thoroughly assess the performance of the proposed 3D and 2D models through precision experiments, in which large deflections and stretching deformations were imposed. The $\mathbf{R}$-based models yielded predictions in excellent agreement with the experimental results for all the cases we considered. As opposed to the $\mathbf{R}$-based models, the $\mathbf{F}$-based 2D and 3D models sometimes failed to predict the deformation of hard-MRE plates. In particular, when non-negligible stretching deformation is present, errors are induced due to the incorrect description of the magnetization of the deformed plate. These disparities were demonstrated to be most significant when the applied field is parallel to the initial magnetization of the plate. We explained the disagreement between the existing $\mathbf{F}$-based theory and the experiments by the magnetically induced pre-stretches/pre-stresses that can be mistakenly predicted in a free/constrained hard-MRE~\citep{Zhao_JMPS2019}. We also showed how the Kirchhoff-Love assumption can artificially correct this discrepancy in the $\mathbf{F}$-based plate model. To the best of our knowledge, our work constitutes the first experimental validation of a rotation-based model for hard-MREs, here using a plate geometry.

The proposed plate model could enable the efficient and systematic exploration of the design space of more complex problems while providing physical insight by analyzing the specialized reduced energy potential. We also highlight that our geometrically nonlinear plate model can be used as a basis to develop simpler plate models by taking additional assumptions on the extents of strains and rotations. When deriving the reduced magnetic potential in Eq.~\eqref{eq:Um_2D_modified}, for simplicity, only the zero-order term was retained. The appropriateness of this approximation is corroborated by the excellent agreement we found between the plate model and experiments on the test cases considered.
However, retaining high-order terms may be needed in problems involving larger stains than those we considered or non-uniform fields. 
We also note that in more complex scenarios, when self-interactions, magnetic hysteresis, or viscoelastic effects of the MRE are non-negligible, the 3D models developed by \cite{mukherjee_explicit_2021} and \cite{rambausek_computational_2022} are to be preferred over ours. Also, when the MRE is not subjected to significant stretching, the $\mathbf{F}$-based model of \cite{Zhao_JMPS2019} remains appropriate, as verified by the comparative results from our first three test cases.

Our work provides a theoretical framework for future studies on the mechanics of hard-magnetic plates. Potential new directions for research include magnetic plate buckling, and the rational design of reconfigurable plate-like structures through magneto-elastic effects. With recent research in MREs gearing towards biomedical applications ~\citep{wang_multi-functional_2022,kim_magnetic_2022}, where precise modeling and closed-loop control are often required, we also believe that our 3D, rotation-based model is timely, providing an accurate and efficient modeling framework for hard-MREs that takes into account the latest developments in the field. 

\section*{Acknowledgement}
The authors are grateful to Matteo Pezzulla for his help in setting up the 2D simulations in COMSOL, building up on his previous work from~\citep{pezzulla_shells_2021}

\appendix
\section{Inverse of stretch tensor $\mathbf{U}^{-1}$}
\label{Appendix_A}

The expression of the inverse of stretch tensor $\mathbf{U}^{-1}$ in terms of deformation gradient $\mathbf{F}$ of a continuum can be found in~\citep{hoger_determination_1984,sawyers_comments_1986,guan-suo_determination_1998}:
\begin{equation}
\begin{split}
\mathbf{U}^{-1}=&\big({I^\mathbf{U}_3}{\Delta^\mathbf{U}}\big)^{-1}\big\{\big[{I^\mathbf{U}_1}\big({I^\mathbf{U}_2}\big)^{2}-{I^\mathbf{U}_3}\big({I^\mathbf{U}_1}\big)^{2}-{I^\mathbf{U}_3}{I^\mathbf{U}_2}\big]\mathbf{I}\\
&-\big[{I^\mathbf{U}_3}+\big({I^\mathbf{U}_1}\big)^{3}-2{I^\mathbf{U}_1}{I^\mathbf{U}_2}\big]\mathbf{C}+I^\mathbf{U}_1\mathbf{C}^2\big\}\,,
\end{split}
\label{eq:inverse_U}
\end{equation}
where $\Delta^\mathbf{U}={I^\mathbf{U}_1}{I^\mathbf{U}_2}-{I^\mathbf{U}_3}$. The three principal invariants of ${\mathbf{U}}$, ${I^\mathbf{U}_1}$, ${I^\mathbf{U}_2}$, and ${I^\mathbf{U}_3}$, are also given in~\citep{hoger_determination_1984,sawyers_comments_1986,guan-suo_determination_1998}
\begin{equation}
\begin{split}
{I^\mathbf{U}_1}=&\sqrt{{I^\mathbf{C}_1}+2{I^\mathbf{U}_2}},\\
{I^\mathbf{U}_2}=&\frac{\sqrt{2\sqrt{I^\mathbf{C}_3}\lambda^3+I^\mathbf{C}_2\lambda^2-I^\mathbf{C}_3}+\sqrt{I^\mathbf{C}_3}}{\lambda},\\
{I^\mathbf{U}_3}=&\sqrt{{I^\mathbf{C}_3}}\,,
\end{split}
\label{eq:IU1_IU2_IU3}
\end{equation}
where ${I^\mathbf{C}_1}$, ${I^\mathbf{C}_2}$, and ${I^\mathbf{C}_3}$ are the three principal invariants of ${\mathbf{C}}=\mathbf{F}^\mathrm{T}\mathbf{F}$, and 
\begin{equation}
\lambda=\frac{1}{\sqrt{3}}\left\{{I^\mathbf{C}_1}+2\left[\left({I^\mathbf{C}_1}\right)^2-3{I^\mathbf{C}_2}\right]^\frac{1}{2}\cos{\left[\frac{1}{3}\cos^{-1}\left(\frac{2\left({I^\mathbf{C}_1}\right)^3-9{I^\mathbf{C}_1}{I^\mathbf{C}_2}+27{I^\mathbf{C}_3}}{2\left({I^\mathbf{C}_1}{I^\mathbf{C}_1}-3{I^\mathbf{C}_2}\right)^\frac{3}{2}}\right)\right]}\right\}^\frac{1}{2}.
\label{eq:lambda}
\end{equation}

\section{Magnetic potential reduced from the \textbf{F}-based 3D theory}
\label{Appendix_B}

To specialize the 3D magnetic potential in Eq.~\eqref{eq:Um_M_exsiting} of the \textbf{F}-based 3D theory for thin plates, we plug the deformation gradient $\mathbf{F}$ from Eq.~\eqref{eq:F_2D} into Eq.~\eqref{eq:Um_M_exsiting}:
\begin{equation}
{U}^\mathrm{m}_\mathbf{F}=-J^{-1}(\hat{\mathbf{F}}\mathbf{M})\cdot\mathbf{B}^\mathrm{a}-\eta_3J^{-1}[(\mathbf{n}_{,\alpha}\otimes\mathbf{e}_{\alpha})\mathbf{M}]\cdot\mathbf{B}^\mathrm{a}\,.
\label{eq:Us_3D_to_2D}
\end{equation}
Equation~\eqref{eq:Us_3D_to_2D} is then integrated over the thickness ($\eta_3\in [-h/2,\,h/2]$) to yield the reduced 2D magnetic potential of a thin plate defined with respect to its mid-surface
\begin{equation}
\hat{U}^\mathrm{m}_\mathbf{F}=-h(\hat{\mathbf{F}}\mathbf{M})\cdot\mathbf{B}^\mathrm{a}\,.
\label{eq:Um_2D_exsiting_appendix}
\end{equation}
To obtain this result, we have assumed that $\mathbf{M}$ and $\mathbf{B}^\mathrm{a}$ are constant along the thickness, and we note that the first-order term vanishes after the integration over $\eta_3 $. The 2D formulation in Eq.~\eqref{eq:Um_2D_exsiting_appendix} is the reduced magnetic potential for the \textbf{F}-based plate model. 

\bibliography{mybibfile}

\end{document}